\documentclass[reprint,aps,superscriptaddress,amsmath,amssymb,longbibliography,nofootinbib]{revtex4-1}
\usepackage{graphicx}
\usepackage{dcolumn}
\usepackage{bm}
\usepackage{color}
\usepackage{xspace}
\usepackage{ulem}
\usepackage{xfrac}
\usepackage[colorlinks=true,urlcolor=blue,citecolor=blue,linkcolor=blue,bookmarks=false,pdfstartview={FitH}]{hyperref}

\def \cm-1{cm$^{-1}$\,}
\footnotetext{These authors contributed equally to this work}
                                                  
\begin{document} 
\title{Symmetry Breaking and Ascending in the Magnetic Kagome Metal FeGe}                            
\author{Shangfei~Wu$^\star$}
\email{wusf@baqis.ac.cn}
\thanks{Present address: Beijing Academy of Quantum Information Sciences, Beijing 100193, China.}
\affiliation{Department of Physics and Astronomy, Rutgers University,
Piscataway, New Jersey 08854, USA}
\author{Mason~L.~Klemm$^\star$}
\affiliation{Department of Physics and Astronomy, Rice University, Houston, Texas 77005, USA}
\author{Jay Shah} 
\affiliation{Department of Chemical Engineering and Materials Science, University of Minnesota, Minnesota 55455, USA}
\author{Ethan T. Ritz} 
\affiliation{Department of Chemical Engineering and Materials Science, University of Minnesota, Minnesota 55455, USA}
\author{Chunruo~Duan}
\affiliation{Department of Physics and Astronomy, Rice University, Houston, Texas 77005, USA}
\author{Xiaokun~Teng}
\affiliation{Department of Physics and Astronomy, Rice University, Houston, Texas 77005, USA}
\author{Bin~Gao}
\affiliation{Department of Physics and Astronomy, Rice University, Houston, Texas 77005, USA}
\author{Feng Ye}
\affiliation{Neutron Scattering Division, Oak Ridge National Laboratory, Oak Ridge, Tennessee 37831, USA}
\author{Masaaki Matsuda}
\affiliation{Neutron Scattering Division, Oak Ridge National Laboratory, Oak Ridge, Tennessee 37831, USA}
\author{Fankang Li}
\affiliation{Neutron Technologies Division, Oak Ridge National Laboratory, Oak Ridge, Tennessee 37831, USA}
\author{Xianghan~Xu}
\affiliation{Department of Physics and Astronomy, Rutgers University, Piscataway, New Jersey 08854, USA}
\author{Ming~Yi}
\affiliation{Department of Physics and Astronomy, Rice University, Houston, Texas 77005, USA}
\author{Turan~Birol} 
\email{tbirol@umn.edu}
\affiliation{Department of Chemical Engineering and Materials Science, University of Minnesota, Minnesota 55455, USA}
\author{Pengcheng~Dai}
\email{pdai@rice.edu}
\affiliation{Department of Physics and Astronomy, Rice University, Houston, Texas 77005, USA}
\author{Girsh~Blumberg} 
\email{girsh@physics.rutgers.edu}
\affiliation{Department of Physics and Astronomy, Rutgers University,
Piscataway, New Jersey 08854, USA}
\affiliation{National Institute of Chemical Physics and Biophysics,
12618 Tallinn, Estonia}
\date{\today}               
                                                                                                                                                                                                              
\begin{abstract}  
     
Spontaneous symmetry-breaking---the phenomenon in which an infinitesimal perturbation can cause the system to break the underlying symmetry---is a cornerstone concept in the understanding of interacting solid-state systems.   
In a typical series of temperature-driven phase transitions, higher-temperature phases are more symmetric due to the stabilizing effect of entropy that becomes dominant as the temperature is increased. However, the opposite is rare but possible when there are multiple degrees of freedom in the system. 
Here, we present such an example of a symmetry-ascending phenomenon upon cooling in a magnetic kagome metal FeGe by utilizing neutron Larmor diffraction and Raman spectroscopy. FeGe has a kagome lattice structure with simple A-type antiferromagnetic order below N\'{e}el temperature $T_N\approx 400$\,K and a charge density wave (CDW) transition at $T_{\text{CDW}}\approx 110$\,K, followed by a spin-canting transition at around 60\,K.
In the paramagnetic state at 460\,K, we confirm that the crystal structure is indeed a hexagonal kagome lattice. On cooling to around $T_{N}$, the crystal structure changes from hexagonal to monoclinic with in-plane lattice distortions on the order of 10$^{-4}$ and the associated splitting of the double-degenerate phonon mode of the pristine kagome lattice. Upon further cooling to $T_{\text{CDW}}$, the kagome lattice shows a small negative thermal expansion, and the crystal structure gradually becomes more symmetric upon further cooling. 
A tendency of increasing the crystalline symmetry upon cooling is unusual; it originates from  
an extremely weak structural instability that coexists and competes with the CDW and magnetic orders. These observations are against the expectations for a simple model with a single order parameter, and hence can only be explained by a Landau free energy expansion that takes into account multiple lattice, charge, and spin degrees of freedom.
Thus, the determination of the crystalline lattice symmetry as well as the unusual spin-lattice coupling is a first step towards understanding the rich electronic and magnetic properties of the system, and it sheds new light on intertwined orders where the lattice degree of freedom is no longer dominant.                                                                                                               
\end{abstract}

                                                                                                                                                                                                                                                                                                                                                                                    
\maketitle

\section{Introduction}                                                                                                                                                                                                                                             Symmetry breaking occurs when a solid changes from one crystalline phase to another at the phase-transition temperature. Landau originally developed a theory of symmetry restrictions on second-order phase transitions in 1937~\cite{Landau1937}. When a crystal structure changes continuously from a highly symmetrical phase to a less symmetrical one, the symmetry group of the low-symmetry phase must be a subgroup of the high-symmetry group~\cite{LandauBook}. Generally, the more symmetrical phase corresponds to higher temperatures, and the less symmetrical phase to lower temperatures. Thus, the symmetry-breaking phenomena are usually detected upon cooling from high temperatures to low temperatures in different types of phase transitions, including structural, incommensurate, magnetic, and liquid crystal systems~\cite{toledano1987landau}. 
Exceptions of the symmetry breaking, namely, symmetry ascending, from the high-temperature phase to the low-temperature phase in a second-order phase transition are reported in 
Rochelle salt crystals (monoclinic to orthorhombic)~\cite{Khan2021} and mixed crystals Tb$_p$Gd$_{1-p}$VO$_4$ (orthorhombic to tetragonal)~\cite{Harley_1974JPC}. These exceptions typically suggest the importance of 
additional interactions such as spin-lattice coupling or cooperative Jahn-Teller interactions to the structural phase transition.  
The symmetry-ascending phenomena are also reported in the first-order-like transitions in iron-based and cuprate superconductors~\cite{Axe_1989PRL,Avci2014,Wang2016PhysRevB,Tidey_2022PRB}.
For example, the parent compound of iron pnictide superconductors order
 antiferromagnetically and ferromagnetically along two Fe-Fe directions of the nearly square lattice to form a stripe antiferromagnetic (AFM) structure~\cite{Dai_2015RMP}.
Since the magnetic structure has the twofold rotational ($C_2$) symmetry, the crystalline lattice must also exhibit a tetragonal-to-orthorhombic ($C_4$ to $C_2$) lattice distortion at temperatures at or above the magnetic ordering temperature $T_N$ to accommodate the low-symmetry magnetic structure~\cite{Dai_2015RMP}. However, when the low-temperature magnetic structure becomes $C_4$ symmetric, as seen in a narrow hole-doped regime of iron pnictides, the lattice symmetry can change from $C_2$ to $C_4$ in a first-order-like fashion due to the formation of the $C_4$ symmetric out-of-plane collinear double-${\bf Q}$ magnetic ordering~\cite{Avci2014,Wang2016PhysRevB,Bohmer2015,Allred2016}. 
This symmetry ascending upon cooling usually suggests that there are several competing interactions near the phase-transition boundary
with similar energy scales~\cite{Christensen_2018PhysRevLett,Christensen_2018PhysRevB}.

\begin{figure*}[!ht] 
\begin{center}
\includegraphics[width=2\columnwidth]{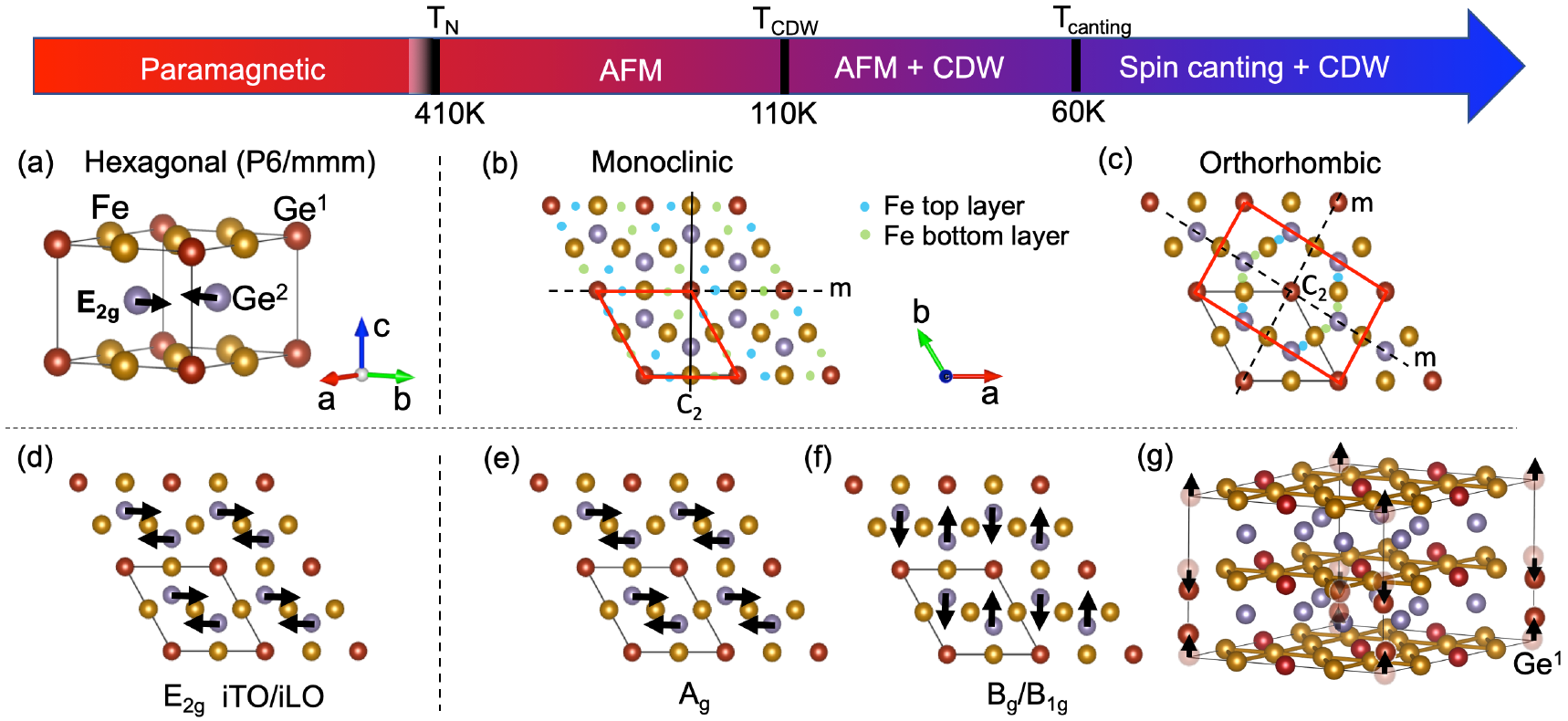}
\end{center}
\caption{\label{Fig1_intro} 
Crystal structure of FeGe in the nonmagnetic phase ($T>T_\text{N}$). 
The Ge atoms in the kagome and honeycomb layers are marked by Ge$^1$ and Ge$^2$, respectively.
The black arrows in panel (a) represent the $E_{2g}$(Ge$^2$) in-plane lattice vibration pattern.
(b) Illustration of the monoclinic AFM unit cell (space group P2$_1$/m) driven by the $A_6^-$ lattice instability shown by the red diamond. The blue and green solid circles represent an example of the distorted Fe atomic positions within the unit cell for the top and bottom layers, respectively. The dashed lines represent the mirror planes. The $C_2$ axis is perpendicular to the threefold axis of the nonmagnetic phase.
(c) Same as panel (b) but for the orthorhombic AFM unit cell (space group Cmcm) driven by the $A_6^-$ lattice instability shown by the red rectangle.
The $C_2$ axis is parallel to the threefold axis of the nonmagnetic phase.
(d) Illustration of the $E_{2g}$ lattice vibration patterns for Ge$^2$ atoms in the honeycomb layers of the hexagonal nonmagnetic phase (top view). This is a twofold-degenerate mode containing an in-plane transverse optical (iTO) mode and an in-plane longitudinal optical (iLO) mode. 
(e) Illustration of the $A_g$ ($A_g$) lattice vibration patterns for Ge$^2$ atoms in the honeycomb layers for the monoclinic (orthorhombic) phases (top view). 
(f) Same as panel (e) but for the $B_g$ ($B_{1g}$) lattice vibration patterns of the monoclinic (orthorhombic) phases. 
The $A_g$ and $B_g$ ($B_{1g}$) modes originate from the splitting of the $E_{2g}$(Ge$^2$) mode of the nonmagnetic phase shown in panel (d). The back arrows in panels (d)-(f) indicate the vibration directions. 
(g) Illustration of the crystal structure of the $2\times2\times2$ CDW phase associated with the $c$-axis dimerization of partial Ge$^1$ atoms in the kagome layer of FeGe based on Refs.~\cite{Miao_arxiv2022_xray,Chen_arxiv2023_STM}.
The black arrows in panel (g) represent the Ge$^1$ displacement directions.}
\end{figure*}

                                                                                                                                                                                                                                                                    The kagome lattice with a corner-shared-triangle network is a fruitful playground to study the exotic electronic orders and symmetry-breaking phenomenon at the interplay between charge, orbital, spin, and lattice degrees of freedom~\cite{Itiro1951,Broholmeaay0668}, including quantum spin liquid, charge density wave (CDW), chiral flux order, nematicity, and superconductivity~\cite{Wang2013PhysRevB,Kiesel2013PhysRevLett,Ortiz2019PhysRevMaterials,FENG_2021SB,Jiang2021NatureMaterial,Denner_arxiv2021,Park_arxiv2021,Tazai_arxiv2021,Xie_neutron_arxiv2021,Chen_arxiv2021,Jiang_arxiv2021_review,III_arxiv2021_usR,Yin2022}.    
Signatures of threefold symmetry breaking have been reported in the CDW phase of the related vanadium-based kagome AV$_3$Sb$_5$ system by scanning tunneling microscopy (STM), $c$-axis magnetoresistance, optical, and x-ray measurements~\cite{WenHH_arxiv2021,Zhao_arxiv2021,Wu_Kerr_arxiv2021,Xu_Kerr_arxiv2022,Kautzsch_arxiv2022_xray}.
Recently, the B35 phase of FeGe with hexagonal kagome lattice structure (P6/mmm, No.~191) [Fig.~\ref{Fig1_intro}(a)]~\cite{Ohoyama1963,Beckman1972,Forsyth,Bernhard_1984,Teng_2022FeGe} has attracted considerable attention due to the interplay amongst different phases. Below  $T_N\approx 400$\,K,
FeGe orders into a collinear A-type AFM structure. The Fe moments within the same basal plane are coupled ferromagnetically with the spin direction parallel to the $c$ axis, while those in adjacent layers are coupled antiferromagnetically~\cite{Bernhard_1984}. Below the spin-canting temperature $T_\text{canting}=60$\,K, the spins form an AFM double-cone structure with a modulated basal-plane moment while the moments are predominantly pointing along the $c$ axis~\cite{Bernhard_1984}. 
At the intermediate temperature, a short-ranged CDW order with a transition temperature $T_{\text{CDW}}\approx 110$ K was discovered by neutron diffraction~\cite{Teng_2022FeGe}, STM~\cite{Yin_2022FeGeSTM}, and angle-resolved photoemission spectroscopy (ARPES) measurements~\cite{Teng_arxiv2022_ARPES}. 
This short-ranged CDW order in FeGe~\cite{Teng_2022FeGe} can be further tuned to be long-ranged by postgrowth annealing treatments~\cite{Miao_arxiv2022_xray,Chen_arxiv2023_STM}.
In particular, CDW order enhances the magnetic-ordered moments, thus establishing a clear coupling between CDW and magnetism~\cite{Teng_2022FeGe}. From subsequent x-ray diffraction measurements, the CDW transition is believed to be associated with the $c$-axis dimerizations of partial Ge$^1$ atoms in the kagome layer of FeGe [Figs.~\ref{Fig1_intro}(a) and~\ref{Fig1_intro}(g)]~\cite{Miao_arxiv2022_xray,Chen_arxiv2023_STM}, but the crystal structure is still refined to be a hexagonal kagome lattice with the threefold rotational ($C_3$) symmetry at all temperatures investigated (from 20\,K to room temperature)~\cite{Chen_arxiv2023_STM,Shi_arxiv2023_Annealing}. Although density functional theory (DFT) phonon calculations have proposed several possible lattice distortion patterns in the CDW phase~\cite{Shao_2022FeGeDFT,Miao_arxiv2022_xray,Zhou_arxiv2022_DFT,Setty_arxiv2022,Ma_arxiv2023_DFT,Wang_arxiv2023_DFT}, inelastic neutron and x-ray scattering measurements did not detect soft acoustic phonon modes at the Brillouin zone boundary across the CDW transition~\cite{Teng_arxiv2022_ARPES,Miao_arxiv2022_xray}.

In this article, we use neutron Larmor diffraction and polarization-resolved Raman spectroscopy to study the temperature-dependent lattice symmetry of FeGe and the associated lattice dynamics.
In the paramagnetic state at 460\,K, we confirm that the crystal structure is indeed a hexagonal kagome lattice [Figs.~\ref{Fig1_intro}(a) and~\ref{Fig1_intro}d)]. On cooling to below $T_{N}$, the $C_3$ symmetry of the hexagonal kagome lattice is broken, which is revealed by 
inequivalent in-plane lattice parameters on the order of $10^{-4}$ and the splitting of a double-degenerate phonon mode [Figs.~\ref{Fig1_intro}(b),~\ref{Fig1_intro}(e), and~\ref{Fig1_intro}(f)]. 
Upon further cooling to a CDW temperature of $T_{\text{CDW}}\approx 110$ K, the kagome lattice shows a small negative thermal expansion, and the crystal structure gradually becomes more symmetric upon further cooling, resulting in a tendency of lattice symmetry ascending below $T_\text{canting}$ [Fig.~\ref{Fig1_intro}(c)]. Since A-type AFM structure is not expected to affect the in-plane kagome lattice structure, our discovery of a series of lattice distortions in the kagome plane as well as the tendency of lattice symmetry ascending upon cooling, suggests an interplay between magnetic order, CDW, and lattice distortion in FeGe.         
Our determination of crystalline lattice symmetry forms the basis for understanding the electronic and magnetic properties of the system. 

\begin{figure*}[!t] 
\begin{center}
\includegraphics[width=2\columnwidth]{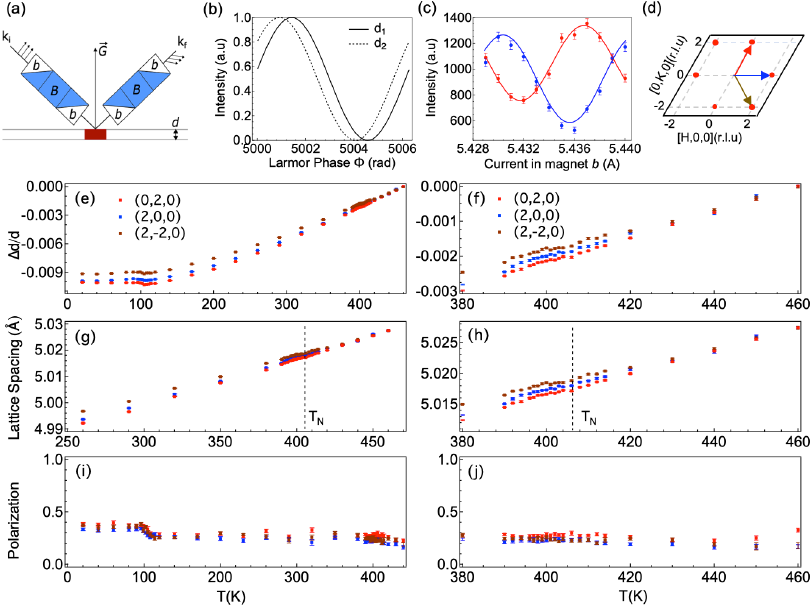}
\end{center}
\caption{\label{Fig2_AFM}
(a) Schematic of the neutron Larmor diffraction instrument setup with Wollaston prisms.
(b) Schematic depicting the Larmor phase difference between two different lattice parameters. 
(c) Example of raw data taken during neutron Larmor diffraction measurements for the (0,~2,~0) peak, where the red and blue points are 170\,K and 20\,K, respectively. The solid lines are sinusoidal fits to the data.
(d) Schematic of the three lattice Bragg peaks (2,~0,~0), (0,~2,~0), and (2,~-2,~0) of the hexagonal phase.
(e) $\Delta d/d$ as a function of temperature measured for (0,~2,~0) (red), (2,~0,~0) (blue), and (2,~-2,~0) (brown) nuclear Bragg peaks from 20\,K to 460\,K.  (f) Same as panel (a) but for 380\,K to 460\,K.
(g) Temperature dependence of the lattice spacing of three theoretically equivalent Bragg peaks for (0,~2,~0) (red), (2,~0,~0) (blue), and (2,~-2,~0) (brown) nuclear Bragg peaks in the [H,~K,~0] plane.
(h) Expanded view of panel (g) near $T_\text{N}$. The dashed lines in panels (g) and (h) mark the $T_\text{N}$. 
Since the error bars are smaller than the symbol, they are not visible in panels (e)-(h).
(i) Neutron polarization as a function of temperature measured for these three nuclear Bragg peaks from 20\,K to 460\,K. (j) Same as panel (c) but for 380\,K to 460\,K.
}
\end{figure*}

\begin{figure*}[!t] 
\begin{center}
\includegraphics[width=2\columnwidth]{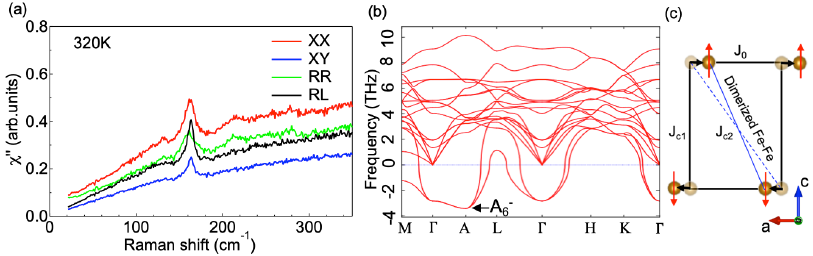}
\end{center}
\caption{\label{Fig_Raman_DFT}
(a) Raman spectra in the $XX$, $XY$, $RR$, and $RL$ scattering geometries at 320\,K. 
(b) DFT phonon dispersion calculation for FeGe in the nonmagnetic phase with $U=2$\,eV and spin not included.
The black arrow indicates the $A_6^-$ lattice instability. The high-symmetry $k$ points follow the conventional notation of the hexagonal Brillouin zone.
(c) Illustration of Fe displacements following the $A_6^-$ lattice instability in a Fe-Fe lattice from the $ac$ plane. Ge atoms are omitted for simplification. 
The black arrows represent the Fe displacement directions while the red arrows  
represent the spin ordering.
In-plane exchange energy is labeled as $J_0$, and the nearest and next-nearest exchange energy along the $c$ axis are labeled as $J_{c1}$ and $J_{c2}$, respectively. The two Fe-Fe connected by $J_{c2}$ with opposite spin orientations may form a dimerlike singlet due to the $A_6^-$ displacements.
}
\end{figure*}                           
                       					             																											
\section{Results}\label{Results} 
          
\textit{AFM phase}\label{AFM}
-- 
To precisely determine the temperature evolution of the lattice symmetry of FeGe across $T_\text{N}$, 
we carried out neutron Larmor diffraction measurements, as shown schematically in Fig.~\ref{Fig2_AFM}(a)~\cite{Lu_2016PhysRevB}.
Neutron Larmor diffraction 
 is a polarized neutron technique developed to increase the resolution of conventional neutron diffractometers by fully taking advantage of the additional spin degree of freedom of the neutron. Inside an applied magnetic field that guides the spin rotation of the traveling neutrons, the neutron spin will experience a motion called Larmor precession around the magnetic-field vectors. The resulting accumulated Larmor phase $\Phi$ is proportional to the magnetic-field intensity $B$, neutron wavelength $\lambda$, and the path length through the magnets $L$. By tilting the field boundaries of the magnets parallel to the crystal plane of interest before and after the sample, the measurement of the lattice spacing $d$ can be linearly translated into the measurement of the Larmor phase as $\Phi=BLd$. The value of $\Phi$ is independent of the beam divergence and the mosaic spread of the sample. For the Larmor diffraction setup at HB-1, magnetic Wollaston prisms~\cite{Li_2014Rev_Sci_Instrum} are utilized such that the physical tilting of the magnetic-field boundaries can be effectively achieved by picking the appropriate combination of the electromagnetic fields shown in Fig.~\ref{Fig2_AFM}(a), i.e., $B$ and $b$. The Larmor phase of the neutron spin is observed by a polarization analyzer to yield $P=\cos(\Phi)$. An example of the measured intensity on the detector has been given in Figs.~\ref{Fig2_AFM}(b) and~\ref{Fig2_AFM}(c) for two different values of the lattice constant.
 By measuring
the relative phase shift between the two intensity oscillations with a high precision ($\Delta \Phi/\Phi \sim 10^{-6}$),
the same resolution is achieved in measuring the change in the lattice constant $\Delta d/d  = \Delta \Phi/\Phi$ for FeGe at the three different Bragg peak positions shown in Fig.~\ref{Fig2_AFM}(d). The temperature dependence of the absolute lattice spacing can be deduced from the measured 
$\Delta d/d$ for FeGe shown in Figs.~\ref{Fig2_AFM}(e) and~\ref{Fig2_AFM}(f).
                                                                                                                                                         
In Fig.~\ref{Fig2_AFM}(g), we show the absolute lattice spacing derived from the three different Bragg peak positions (2,~0,~0), (0,~2,~0), and (2,~-2,~0). Above 430\,K, the temperature dependence of the three lattice spacings is identical. This finding is consistent with an ideal kagome lattice with threefold lattice rotational symmetry. 
Below $T_\text{N}$, the temperature dependence of the lattice spacing along the (2,~0,~0), (0,~2,~0), and (2,~-2,~0) directions deviates, indicating threefold symmetry breaking. The clear three lattice spacing below $T_\text{N}$ indicates that the AFM phase is monoclinic (Appendix~\ref{Bragg_peaks}). 
It also suggests that the sample is in a single domain. As discussed in Refs.~\cite{Lu_2016PhysRevB,Wang_2018NC}, neutron Larmor diffraction can measure the broadening of the Bragg peaks due to lattice distortion formation of twinned domains [see Fig.~1(c) of Ref.~\cite{Wang_2018NC} and Fig.~4 of Ref.~\cite{Lu_2016PhysRevB}]. In the case of NaFeAs (Ref.~\cite{Wang_2018NC}), the samples are twinned, and we can precisely measure the lattice parameter change above and below twinning. In fact, twinning of the crystal means that multiple Bragg peaks with different lattice parameters occur at approximately the same position in reciprocal space but become broader below the formation of twin domains. Since the lattice parameter correlates linearly with the neutron spin's Larmor phase, an expansion or broadening of the nuclear Bragg peak corresponds to a similar change or dispersion in the neutron spin's Larmor phase. Consequently, any distortion or broadening in the lattice structure and formation of twin domains would manifest as variations in the neutron beam's polarization. This is clearly seen below the N\'{e}el temperature of YBa$_2$Cu$_3$O$_6$, the AFM-ordered parent compound of cuprate superconductors, in the neutron Larmor diffraction experiment~\cite{Nafradi_2016PRL}.  For FeGe, the observed flatness of neutron polarization across $T_\text{N}$ shown in Figs.~\ref{Fig2_AFM}(i) and~\ref{Fig2_AFM}(j) signifies the absence of twinning or broadening during this transition.      
Surprisingly, the deviation of the three lattice spacing appears slightly above $T_N$, which extends about 20\,K above $T_N$, as shown in Fig.~\ref{Fig2_AFM}(h). 
The deviation might be due to the fluctuations of the order parameter above $T_\text{N}$, suggesting a coupling between the magnetism and lattice.
The lattice-spacing difference between (2,~0,~0) and (2,~-2,~0) at room temperature is about $(d_{2\bar{2}0}-d_{200})/(d_{2\bar{2}0} + d_{200}) \approx 3 \times 10^{-4}$. For comparison, the orthorhombic lattice distortion in NaFeAs pnictide is 
$(a_o- b_o)/(a_o+b_o) \approx 1.7 \times 10^{-3}$ (where $a_o$ and $b_o$ are orthorhombic lattice parameters below the structure phase-transition temperature of 58\,K), about 5 times larger~\cite{Wang_2018NC}. 
Therefore, the AFM phase of FeGe is not an ideal kagome lattice but exhibits a weak lattice distortion around $T_\text{N}$.
                               
Furthermore, nuclear structure factor analysis indicates that the structure factor contribution from the Ge sublattice cancels out in the undistorted hexagonal phase for $(2,0,0)$ and its equivalent Bragg peaks.
For the distorted lattice, the Ge and Fe atoms' structure factor contributions to the now inequivalent $(2,0,0)$, $(0, 2, 0)$ and $(2, -2, 0)$ Bragg peaks are approximately destructive and constructive, respectively (Appendix~\ref{Structure_factor_analysis}).
Thus, the distinct three lattice spacing at (2,~0,~0), (0,~2,~0), and (2,~-2~0) Bragg peaks suggests the presence of the symmetry-breaking in-plane Fe-sublattice distortions in the kagome plane below $T_\text{N}$.    
                                                                                                                                                                                                                                                                                                                                                                 
The threefold symmetry breaking is also revealed in Raman spectroscopy.                                                             
Above $T_\text{N}$, FeGe has the space group P6/mmm (\#191) with three formula units in the unit cell. Fe ions occupy the Wyckoff site $3f$, whereas Ge ions occupy $1a$ (in-plane) and $2d$ (apical) sites. 
The $\Gamma$-point phonon modes transform as 
$\Gamma_\text{total} = 3A_{2u} \oplus B_{2g} \oplus B_{1u} \oplus B_{2u} \oplus E_{2u} \oplus E_{2g} \oplus 4E_{1u}$,       
where there is only one two-fold degenerate Raman-active mode $\Gamma_{\text{Raman}}= E_{2g}$(Ge$^2$)~[Ge$^2$ are the apical Ge ions, which form the honeycomb layer shown in Fig.~\ref{Fig1_intro}(a)] (Appendix~\ref{GroupTheory}). 
In Fig.~\ref{Fig_Raman_DFT}(a), we show the Raman spectra of FeGe at room temperature in the AFM phase.
One main phonon at around 160-162~\cm-1 is detected in all four ($XX$, $XY$, $RR$, and $RL$) scattering geometries (Appendix~\ref{Methods}). 
However, the peak positions in the $RR$ spectrum are at a slightly lower frequency than in the $RL$ spectrum. 
One mode at 160\,\cm-1 is observed in the $RR$ scattering geometry while a mode at 162\,\cm-1 is observed in the $RL$ scattering geometry.
The proximity in energy between these two modes is attributed to their common origin from the splitting of the degenerate Raman-active $E_{2g}$(Ge$^2$) mode of the honeycomb layer in the nonmagnetic phase. The detection of these two split modes indicates threefold rotational symmetry breaking in the AFM phase. The splitting of the $E_{2g}$(Ge$^2$) mode is only 2\,\cm-1 at room temperature, consistent with the rather small lattice distortion observed in the neutron Larmor diffraction measurements shown in Fig.~\ref{Fig2_AFM}(g).
               
\begin{figure}[!t] 
\begin{center}
\includegraphics[width=\columnwidth]{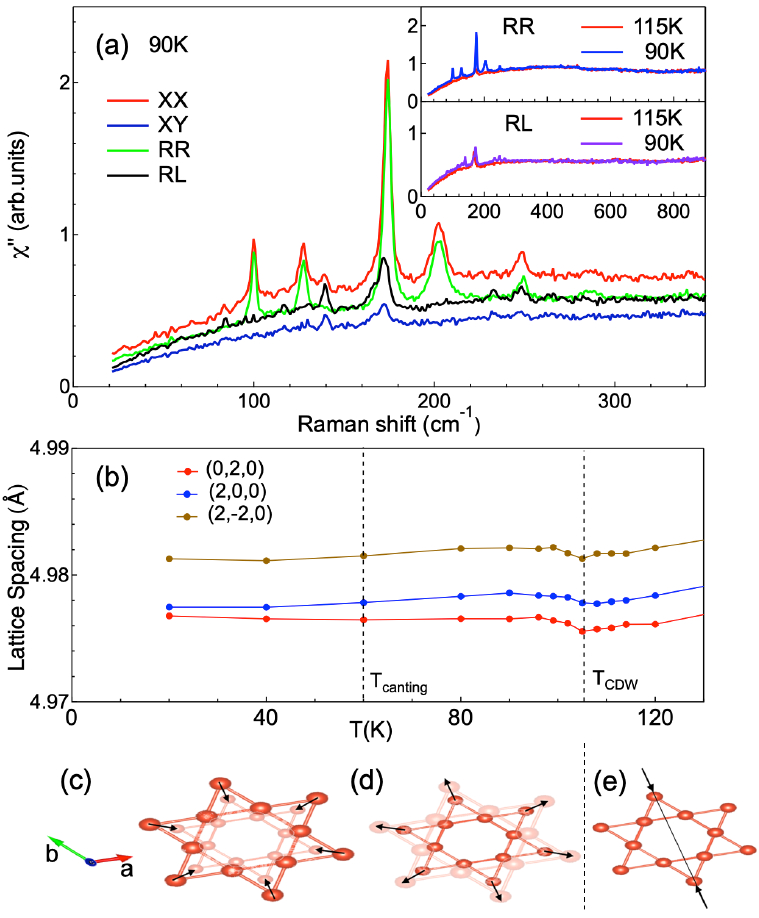}
\end{center}
\caption{\label{Fig3_CDW}
(a) Raman spectra in the $XX$, $XY$, $RR$, and $RL$ scattering geometries at 90\,K. The inset of panel (a) shows the Raman response in the $RR$ (top panel) and $RL$ (bottom panel) scattering geometries in an extended energy range up to 900\,\cm-1 at 115\,K and 90\,K.
(b) Temperature dependence of the $d$-spacing of three Bragg peaks for (0,~2,~0) (red), (2,~0,~0) (blue), and (2,~-2,~0) (brown) nuclear Bragg peaks in the [H,~K,~0] plane below 130\,K. The dashed lines in panel (b) mark $T_\text{CDW}$ and $T_\text{canting}$.
(c)--(e) Schematic of lattice distortion observed using neutron Larmor diffraction for temperatures above $T_\text{CDW}$ and below $T_\text{CDW}$. Above $T_\text{CDW}$, panel (e) illustrates the monoclinic lattice distortion. Below $T_\text{CDW}$, the lattice first displays a negative thermal expansion, and the unit-cell volume becomes larger, as shown in panel (d). Upon further cooling below $T_\text{CDW}$, the unit-cell volume then becomes smaller again, as shown in panel (c). The kagome lattices shown in panels (c)--(e) are depicted in real space with exaggerated distortions for visual clarity. 
}
\end{figure}

In order to explore the origin of the threefold symmetry breaking in FeGe, we performed first-principles (density functional theory) lattice response calculations to search for the lattice instabilities. 
The phonon dispersion displays no instabilities when the (collinear) AFM order is imposed in the calculation, which is consistent with Refs.~\cite{Shao_2022FeGeDFT,Miao_arxiv2022_xray,Ma_arxiv2023_DFT,Wang_arxiv2023_DFT}. This finding may be because DFT, being a static mean field theory, typically overestimates the ordered magnetic moments and hence does not correctly capture certain electronic features such as the orbital occupations. When we repeat the DFT calculations without a magnetic order, we find strong instabilities at the $\Gamma$ and $A(0,~0,~0.5)$ points of the Brillouin zone.                
In Fig.~\ref{Fig_Raman_DFT}(b), we show these DFT phonon dispersions obtained for the nonmagnetic FeGe with an added on-site Coulomb interaction $U=2$\,eV. A complete branch of the twofold degenerate phonon dispersion that includes $\Gamma_6^-$ and $A_6^-$ modes has imaginary frequencies, indicating lattice instabilities. The largest imaginary frequency is from the $A_6^-$ mode, and this mode remains unstable in many different configurations (Appendix~\ref{DFT_robustness}). Both of the unstable modes $A_6^-$ and $\Gamma_6^-$ can lead to either orthorhombic or monoclinic structures: $A_6^-$ can drive a transition to space groups Cmcm (\#63, point group $D_{2h}$) or P2$_1$/m (\#11, point group $C_{2h}$), and $\Gamma_6^-$ can lead to space groups Amm2 (\#38) or Pm (\#6). Details of the group-subgroup analysis are presented in Appendix~\ref{Group_subgroup_analysis}. 
The main difference between $A_6^-$ and $\Gamma_6^-$ is that $A_6^-$ leads to a doubling of the unit cell along the $c$-axis direction while $\Gamma_6^-$ does not because $A_6^-$ has the same wave vector as the collinear A-type AFM order. 
While we explicitly focus on the $A_6^-$ mode in the rest of this paper, a similar argument also applies to the $\Gamma_6^-$ mode, and we cannot distinguish these different structures driven by $A_6^-$ or $\Gamma_6^-$ instabilities within the resolution of the present Raman data (Appendix~\ref{Group_subgroup_analysis}).
The monoclinic and  orthorhombic structures below $T_N$ that the $A_6^-$ mode leads to are illustrated in Figs.~\ref{Fig1_intro}(b) and~\ref{Fig1_intro}(c), respectively.
                                                                                                                                                                                                                                                                                                                                                                                                                                                                                                                                                                                                    
The symmetry breaking at $T_\text{N}$ driven by the $A_6^-$ lattice instability mainly involves displacements of Fe or Ge$^1$ in the kagome layer. The resulting displacements have opposite directions for the top and bottom kagome layers (Appendix~\ref{A6appendix}). Thus, the Ge$^2$ atoms in the honeycomb layer experience an anisotropy of the local crystal electrical field [Fig.~\ref{Fig1_intro}(b)]. As a consequence, the fundamental $E_{2g}$(Ge$^2$) mode from the honeycomb layers splits into $A_g \oplus B_g$ ($C_{2h}$ point group) for the monoclinic AFM phase. This explains the two modes we observed at room temperature shown in Fig.~\ref{Fig_Raman_DFT}(a). The mode observed in the $RR$ scattering geometry is attributed to the $A_g$ mode while the one observed in the $RL$ scattering geometry is attributed to the $B_g$ mode (Appendix~\ref{Raman_tensor_analysis}). 
Illustrations of the $A_g$ and $B_g$ lattice vibration patterns for the Ge$^2$ atoms are shown in Figs.~\ref{Fig1_intro}(e) and~\ref{Fig1_intro}(f), respectively.

\textit{CDW phase} \label{CDW_phase}                                                                      
 --                                                                                                                                                                                                                                                  Below 110\,K, FeGe undergoes a CDW transition~\cite{Teng_2022FeGe}. 
Recent x-ray studies show that the crystal structure of the $2\times2\times2$ CDW phase is associated with the $c$-axis dimerization of Ge$^1$ atoms in the kagome layer of FeGe [Fig.~\ref{Fig1_intro}(g)] based on Refs.~\cite{Miao_arxiv2022_xray,Chen_arxiv2023_STM}.
New phonon modes are expected to appear in the CDW state due to Brillouin zone folding~\cite{Devereaux2007RMP}.
Only those modes modulating the ionic deviation above $T_\text{CDW}$ with a large amplitude can obtain noticeable Raman intensity below $T_\text{CDW}$ and thus can be detected in the Raman spectra~\cite{Klein1982PhysRevB,Holy1976PRL,NAGAOSA1982809}.                                                                                                                                                   
In Fig.~\ref{Fig3_CDW}(a), we show several new phonon modes appearing in all four scattering geometries at 90\,K in the CDW state. 
Specifically, four additional modes are detected in the $RR$ scattering geometry, and eight additional modes are observed in the $RL$ scattering geometry below $T_{\text{CDW}}$. These modes are the amplitude modes of the CDW order parameter.                
These new phonon peak positions are summarized in Appendix~\ref{Phonon_peak_position_CDW_phase}. 
                                                                                                                                                                                                                                                           
In the inset of Fig.~\ref{Fig3_CDW}(a), we show the Raman response in both $RR$ and $RL$ scattering geometries up to 900\,\cm-1 for above and below $T_\text{CDW}$. The CDW gap-opening signatures, namely the suppression of the low-energy spectra weight and the enhancement of the spectra weight close to $2\Delta\approx 50$\,meV determined by STM~\cite{Yin_2022FeGeSTM,Chen_arxiv2023_STM} and ARPES~\cite{Teng_arxiv2022_ARPES}, are not observed in the Raman response. The absence of CDW gap-opening signatures may be due to the multiband effects in FeGe, similar to the AV$_3$Sb$_5$ system~\cite{SFWU_PhysRevB2022}. 
                                                                                                                                            
\begin{figure*}[!t] 
\begin{center}
\includegraphics[width=2\columnwidth]{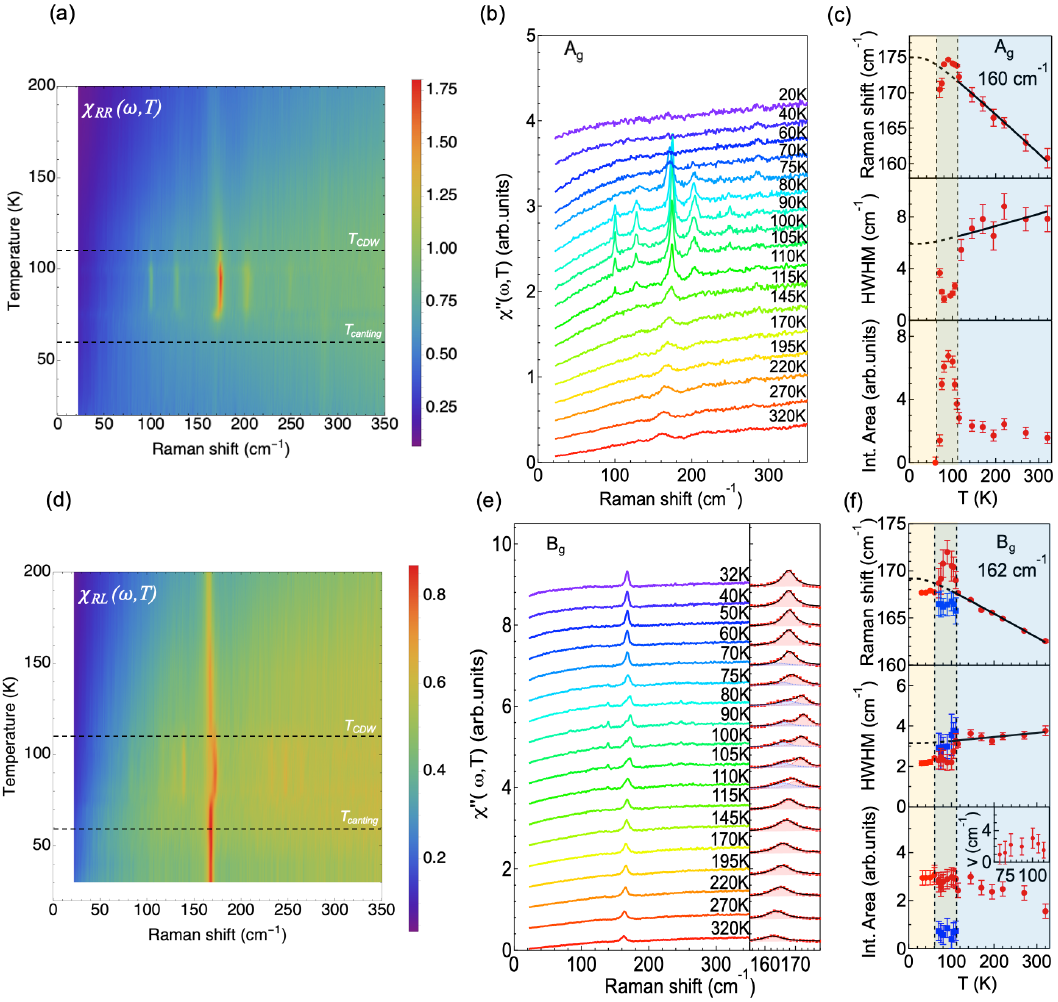}
\end{center}
\caption{\label{Fig4_T_dependence} 
(a) Color plot of the $T$ dependence of Raman response in the $RR$ scattering geometry for FeGe. 
(b) Corresponding Raman spectra of panel (a). 
(c) $T$ dependence of the peak position, half width at half maximum (HWHM), and integrated intensity for the $A_{g}$ phonon mode at 160~\cm-1.
The error bars represent one standard deviation.
The $T$ dependence of phononic frequency and HWHM are fitted by an anharmonic phonon decay model (Appendix~\ref{Anharmonic_decay_model}). 
The dashed lines in panels (a) and (c) represent $T_\text{CDW}$ and $T_\text{canting}$.
(d) Color plot of the $T$ dependence of Raman response in the $RL$ scattering geometry for FeGe.
(e) Corresponding Raman spectra of panel (d). The right part of panel (e) shows a zoom-in of the $T$ dependence of the $B_{g}$ phonon mode at around 162~\cm-1 and its low-energy shoulder peak. The black solid lines are the fitting curve for the total coupled response. The red and blue shaded areas represent the two bare modes. 
(f) $T$ dependence of the peak position, HWHM, and integrated intensity for the bare $B_{g}$ phonon mode at around 162~\cm-1(red) and the bare shoulder peak (blue). The inset in the bottom panel of (f) is the fitted interaction strength between the bare $B_{g}$ phonon mode at 162~\cm-1 and the bare shoulder peak.
The error bars represent one standard deviation.
The $T$ dependence of phononic frequency and HWHM are fitted by an anharmonic phonon decay model (Appendix~\ref{Anharmonic_decay_model}). 
}
\end{figure*}                                                                      
                                                                                                                                              
After establishing the Raman spectroscopic signature of the CDW state, we present the lattice response data below $T_\text{CDW}$.
In Fig.~\ref{Fig3_CDW}(b), we show the temperature dependence of the three lattice spacings upon cooling across $T_\text{CDW}$. In the AFM phase above $T_\text{CDW}$, the unit cell is monoclinic, as illustrated in Fig.~\ref{Fig3_CDW}(e).
The three lattice-spacings decrease upon cooling, following the normal thermal expansion rule.
Below $T_\text{CDW}$, the three lattice spacings display anomalies and increase upon cooling, indicating a small negative thermal expansion, as illustrated in Fig.~\ref{Fig3_CDW}(d). Upon further cooling, the three lattice spacings decrease again, and the unit cell volume become smaller, as we illustrate in Fig.~\ref{Fig3_CDW}(c).
To our surprise, the difference between the lattice spacing for the lattice Bragg peaks (0,~2,~0) and (2,~0,~0) become smaller and smaller, and tend to be almost the same size at the lowest temperature 20\,K. This finding indicates that the monoclinic lattice distortion is weakened and that the lattice tends to be orthorhombic at low temperatures (Appendix~\ref{Bragg_peaks}), signifying that the lattice symmetry tends to ascend at low temperatures.                                                                                                                                                             
                                                                                                                                                                                                                                         
The weakening of the monoclinic lattice distortion below $T_\text{CDW}$ is also seen from the temperature dependence of the Raman modes.                                                                                    
In Figs.~\ref{Fig4_T_dependence}(a) and~\ref{Fig4_T_dependence}(b), we present the $T$ dependence of the phonon modes in the $RR$ scattering geometries. 
The phonon modes at 100\,\cm-1, 127\,\cm-1, 203\,\cm-1, and 249\,\cm-1 appear abruptly below $T_{\text{CDW}}$. In contrast, the $A_g$ phonon at 160\,\cm-1 continuously evolves into the CDW phase. As shown in Fig.~\ref{Fig4_T_dependence}(c), the mode hardens and sharpens upon cooling and experiences additional hardening and sharpening below $T_\text{CDW}$. However, it softens and broadens when approaching $T_\text{canting}$ and finally disappears below $T_\text{canting}$. The broadening of the Raman modes when approaching $T_\text{canting}$ is also found for the other three modes at 100~\cm-1, 127~\cm-1, and 203~\cm-1 in the $RR$ scattering geometry of the CDW phase~(Appendix~\ref{Fitting_parameters_for_CDW_modes}).
Remarkably, the integrated intensity for the $A_g$ phonon at the 160\,\cm-1 mode is enhanced 3.5 times below $T_\text{CDW}$ but decreases to zero close to $T_\text{canting}$.                                                                                                                                                                                                                                                                                                                                                                                                                                                                                                                                                                                                                                                                                                                                                           In Figs.~\ref{Fig4_T_dependence}(d) and~\ref{Fig4_T_dependence}(e), we present the $T$ dependence of the phonon modes in the $RL$ scattering geometries. The $B_g$ phonon at 162\,\cm-1 persists from room temperature down to 32\,K. In contrast, several new modes appear in the CDW phase and disappear below $T_{\text{canting}}$.
In particular, a shoulder peak develops on the lower-energy side of this phonon below $T_\text{CDW}$, as is shown in a zoom-in plot on the right part of Fig.~\ref{Fig4_T_dependence}(e). 
The mode at 162\,\cm-1 and the shoulder mode can be described by a coupled two-Lorentzian-phonon model on a linear background (Appendix~\ref{Fitting_model_for_B1g_modes}), as shown in Fig.~\ref{Fig4_T_dependence}(e).             
In Fig.~\ref{Fig4_T_dependence}(f), the $B_g$ mode at 162\,\cm-1 hardens and sharpens upon cooling and experiences additional hardening and sharpening below $T_\text{CDW}$. It softens sharply when approaching $T_\text{canting}$ and barely change below $T_{\text{canting}}$. 
The weakening and disappearance of the $A_{g}$ mode at 160\,\cm-1 in the $RR$ scattering geometry and a single peak at 162\,\cm-1recovering in the $RL$ scattering geometry below $T_\text{canting}$ indicate that the monoclinic lattice distortion is weakened at lower temperatures. It might be too weak to give rise to any noticeable Raman intensity in the $RR$ scattering geometry.
                                                                                                                                                                                                                                                                                                                                                                                                                                                  
\section{Discussion and Conclusion}\label{Discussions1} 
                                                                                                                                                                                                                                                                         
The $A_6^-$ lattice instability mainly involves in-plane Fe or Ge$^1$ displacements in the kagome layer, leading to the threefold symmetry breaking. The in-plane Fe displacement was revealed in a recent x-ray refinement study, though it is not an A$_6^-$ type yet~\cite{Shi_arxiv2023_Annealing}.
The degenerate fundamental $E_{2g}$(Ge$^2$) mode in the nonmagnetic phase splits into two modes below $T_\text{N}$ due to the symmetry breaking.
However, we do not detect any additional noticeable phonon intensity related to the Fe or Ge$^1$ vibration modes in the Raman spectra at room temperature, as shown in Fig.~\ref{Fig_Raman_DFT}(a). We note that the x-ray scattering measurement did not detect the dimerization lattice Bragg peak at $(H,~K,~L+0.5)$ ($H, K, L$ are integers) between $T_\text{N}$ and $T_\text{CDW}$~\cite{Miao_arxiv2022_xray,Shi_arxiv2023_Annealing}.
This finding might be due to the small amplitude of the in-plane Fe and Ge$^1$ displacements, as the monoclinic lattice distortion is close to 0.03\% at room temperature [Fig.~\ref{Fig2_AFM}(g)], or the fact that the in-plane $A_6^-$ displacements do not modulate the distance between the layers.

The neutron Larmor diffraction results [Fig.~\ref{Fig2_AFM}(g)] indicate in-plane lattice distortion in the kagome plane in the AFM ordered phase. Furthermore, recent inelastic neutron scattering results show that the spin-wave dispersion along the $c$-axis direction displays a spin gap of about 1\,meV at room temperature at $(0,0,0.5)$~\cite{Chen_arxiv2023_neutron}. The AFM order below $T_\text{N}$, accompanied by in-plane lattice distortion and a spin-gap opening in the magnetic excitation spectrum, suggests that the AFM transition at $T_N$ is a likely spin-Peierls-like transition driven by spin-lattice coupling~\cite{peierls_1955book}. Indeed, as we show in Fig.~\ref{Fig_Raman_DFT}(c), the $A_6^-$ lattice instability displaces the Fe atoms in opposite directions for two adjacent kagome planes, thus modulating the interlayer Fe-Fe distance. In particular, the two Fe-Fe distance connected by next-nearest exchange energy $J_{c2}$ becomes shorter and may possibly form a dimerlike singlet, creating the spin gap and spontaneously aligning the moment along the $c$-axis direction. The existence of such a spin gap could lead to a decrease in magnetic free energy that outweighs the increase in lattice free energy due to the distortion~\cite{Bulaevskii_1978,Bray_1983book}. Thus, a compete understanding of the AFM transition at $T_N$ needs to treat the $A_6^-$ lattice displacements, magnetic interactions, and spin-lattice coupling on equal footing.

The neutron Larmor diffraction results [Fig.~\ref{Fig2_AFM}(g)] reveal in-plane lattice distortions $(d_{200}-d_{020})/(d_{200}+d_{020}) \sim 0.02\%$ and $ (d_{2\bar{2}0}-d_{200})/(d_{2\bar{2}0}+d_{200}) \sim 0.04\%$ in the kagome plane of the AFM ordered phase at around 110-115\,K, with an average lattice distortion of about 0.03\%.
The corresponding split-phonon anisotropy for the $A_g$ and $B_g$ modes is $(\omega_{A_g}-\omega_{B_g})/(\omega_{A_g}+\omega_{B_g})= (172-168)/(172+168)=1.2$\%, which is about 40 times that of the lattice anisotropy. In comparison, the AFM phase of EuFe$_2$As$_2$ has a split-phonon anisotropy of about 4\% and a lattice distortion of about 0.55\% at 30\,K~\cite{Zhang_2016PhysRevB}. The ratio between split-phonon anisotropy and the lattice distortion is about 7 for EuFe$_2$As$_2$. The larger ratio between split-phonon anisotropy and the lattice distortion in FeGe suggests additional interactions such as spin-lattice (spin-phonon) coupling play an important role in creating a large split-phonon anisotropy with a tiny lattice distortion~\cite{Chauviere_2009PhysRevB}.

The CDW phase evolves from the monoclinic AFM phase.  Since the monoclinic lattice distortion is about 0.03\% at room temperature, we can thus approximately regard the monoclinic AFM phase as a hexagonal lattice with a weak monoclinic lattice distortion as a perturbation. In this case, the CDW phase could be approximately driven by three $L$-point lattice instabilities of the hexagonal lattice, leading to a $2\times2\times2$ reconstruction of the AFM phase. 
Below $T_\text{CDW}$, the CDW order parameter coexists with the monoclinic lattice distortion, persisting to the lowest temperature, and it does not show any noticeable changes upon cooling across $T_\text{canting}$~\cite{Teng_2022FeGe}. From Fig.~\ref{Fig3_CDW}(b), the monoclinic lattice distortion is already weakened well above $T_\text{canting}$ in the CDW phase.
Similarly from Fig.~\ref{Fig4_T_dependence}(c), the $A_g$ mode at 160\,\cm-1 starts to weaken and broaden below 90\,K, which is also above $T_\text{canting}$ in the CDW phase.  
The temperature dependence of the CDW order parameter and the monoclinic lattice distortion suggests that the $A_6^-$ monoclinic lattice distortion is weakened by coupling to the CDW order parameter.
We note that the spin-canting order parameter could also be coupled to the $A_6^-$ order parameter and help weaken the monoclinic lattice distortion. Thus, both coupling mechanisms could contribute to weakening the monoclinic lattice distortion, leading to the lattice symmetry ascending at lower temperatures. One possible scenario to explain the Raman data below $T_\text{canting}$ is that the lattice symmetry recovers to the hexagonal symmetry, which prohibits the splitting between the $A_g$ mode at 160\,\cm-1 and the $B_g$ mode at 162\,\cm-1. 
This scenario differs from the lattice constant data from Larmor diffraction, where FeGe tends to become orthorhombic when approaching 10\,K. We note that the Raman phonon intensity in FeGe is generally 2 orders of magnitude weaker than that in the AV$_3$Sb$_5$ compound~\cite{SFWU_PhysRevB2022}. The weak Raman signal makes the study of Raman phonons rather challenging in FeGe. The origin of these sharp changes in the Raman data close to $T_\text{canting}$ is an open question and calls for future investigation.

The interplay between the $A_6^-$ monoclinic lattice distortion, CDW, and spin-canting order parameters can be captured by a phenomenological free energy model constructed in the nonmagnetic hexagonal P6/mmm phase with the highest symmetry. The free energy of FeGe in terms of the unstable $A_6^-$ mode (two-dimensional, with components $A_i$, $i = 1, 2$), the $L$-point bond-order CDW modes (either $L_2^-$ or $L_1^+$ depending on the origin choice, with components $L_i$, $i = 1, 2, 3$), and the spin-canting order parameter $S$
is the collection of all polynomials that are invariant under all symmetry operations of the parent space group. It can be written up to fourth order as
\begin{equation}\label{free_energy111}
\begin{aligned}   
\mathcal{F}=\mathcal{F}_A+\mathcal{F}_L+\mathcal{F}_{A L}+\mathcal{F}_S+\mathcal{F}_{A S}
\end{aligned}
\end{equation}  
\begin{equation}\label{free_energy222}
\begin{aligned} 
\mathcal{F}_A=\alpha_A A^2+\beta_A A^4
\end{aligned}
\end{equation}  
\begin{equation}\label{free_energy333}
\begin{aligned}           
\mathcal{F}_L=\alpha_L L^2+\beta_L L^4+\lambda_L (L_1^2 L_2^2+L_2^2 L_3^2+L_3^2 L_1^2)
\end{aligned}
\end{equation}
 \begin{equation}\label{free_energy444}
\begin{aligned}
\mathcal{F}_{A L}
&=\lambda_{A L} A^2 L^2+\gamma_{A L}[A_1^2 (\frac{5}{6} L_1^2+\frac{2}{6} L_2^2+\frac{5}{6} L_3^2)\\
&+A_2^2(\frac{1}{2} L_1^2+L_2^2+\frac{1}{2} L_3^2)+\frac{1}{\sqrt{3}} A_1 A_2(-L_1^2+L_3^2)].
\end{aligned}
\end{equation} 
\begin{equation}
	\mathcal{F_S}= \alpha_S S^2 + \beta_S S^4 
\end{equation}
\begin{equation}
	\mathcal{F_{AS}}= \lambda_{AS} A^2 S^2 
\end{equation}
There is no third-order coupling between $A_i$ and $L_i$. Even though the fourth-order couplings between the two order parameters $A$ and $L$ are not isotropic, the generic biquadratic coupling term $\lambda_{AL}$ is sufficient to explain the competition between $A$ and $L$ order parameters. The second-order coefficient of $A$ is renormalized when $L$ is nonzero according to $F_A = (\alpha_A + \lambda_{AL} L^2)  A^2 + \beta_A A^4$. If $\lambda_{AL}>0$, the CDW order parameter $L_2^-$(or $L_1^+$) would suppress $A_6^-$ and make the monoclinicity disappear eventually.
                                     
Since the spin-canting order parameter $S$ transforms as a time-reversal odd irreducible representation at the incommensurate wave vector, it can couple with the $A$ order parameter only at the biquadratic level or higher. Since the monoclinic distortion is small, its energy scale can be comparable to that of the spin-canting order in FeGe; in other words, $|\alpha_A| \sim |\lambda_{AS}S^2|$. This finding would explain that, as the spins are canted at low temperature, its monoclinicity is weakened because the additional competition between $S$ and $A$ would help suppress the $A$ order parameter.

\begin{figure}[t] 
\begin{center}
\includegraphics[width=\columnwidth]{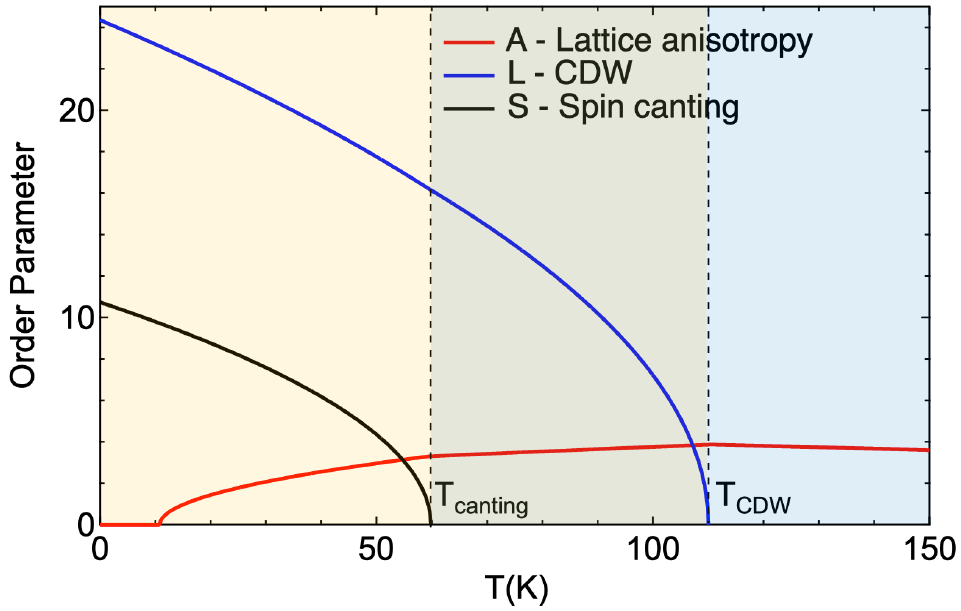}
\end{center}
\caption{\label{Fig5_free_energy} 
Illustration of the Landau free energy model in Eq.~(\ref{free_energy111}). The parameters used in the model are $\alpha_{A_0}=0.1$, $\beta_A=1$, $\alpha_{L_0}=0.1$, $\beta_L=0.01$, $\lambda_{AL} =0.05$, $\lambda_{L} =0$, $\gamma_{AL} =0$, $\alpha_{S_0}=0.2$, $\beta_S=0.06$, $\lambda_{AS} =0.14$, $T'_\text{N}=410$,  $T'_\text{CDW}=117.5$, and $T'_\text{canting}=67.5$. The red, blue, and black solid lines represent the $T$ dependence of the order parameters $A$, $L$, and $S$, respectively. The two dashed lines represent $T_\text{CDW}$ and $T_\text{canting}$.
The blue and yellow shaded areas represent the AFM phase ($T_\text{canting}<T<T_\text{N}$) and the CDW phase ($T<T_\text{CDW}$), respectively. 
}
\end{figure}

Setting $\alpha_{A}(T)=\alpha_{A_0}(T-T'_\text{N})$, $\alpha_L(T)=\alpha_{L_0}(T-T'_\text{CDW})$, and $\alpha_S(T)=\alpha_{S_0}(T-T'_\text{canting})$, where  $T'_\text{N}$, $T'_\text{CDW}$, and $T'_\text{canting}$ are the phase-transition temperatures without the coupling terms, and minimizing the free energy in Eq.~(\ref{free_energy111}) with respect to $A$, $L$, and $S$, we obtain the solution of $T$ dependence of the $A(T)$, $L(T)$, and $S(T)$. While it is not possible to determine the parameters of the free energy precisely, we could choose a combination that reproduces the experimental observations. 
There are other possibilities with different contributions from the CDW or incommensurate magnetic order parameter that suppress the monoclinic lattice distortions at low temperatures. For illustration purposes, we choose the parameters to reproduce the transition temperatures, and the fact that the lattice anisotropy (here represented by the $A$ order parameter) is suppressed by both the CDW and the incommensurate magnetic order parameters.
As we show in Fig.~\ref{Fig5_free_energy}, positive coupling constants $\lambda_{A L}$ and $\lambda_{A S}$ weaken the monoclinic distortion and make it disappear below $T_\text{canting}$.
Other possibilities describing the interplay between $A$ and $L$ order parameters
may involve the $\gamma_{A L}$ term, which prefers a certain direction of $A$ order parameters depending on the direction of $L$ order parameters.

The interplay between the $A_6^-$ monoclinic lattice distortion, CDW, and magnetic order originates from the fact that they are competing orders with similar energy scales. This case is similar to the $C_4$ reentrance phase observed in the hole-doped iron pnictides, where the formation of an out-of-plane collinear double-${\bf Q}$ magnetic ordering~\cite{Avci2014,Wang2016PhysRevB,Meier2018,Allred2016,Bohmer2015} leads to the restoration of lattice symmetry from orthorhombic to tetragonal upon cooling due to competing orders~\cite{Christensen_2018PhysRevLett,Christensen_2018PhysRevB}. The difference is that the weakening of the $A_6^-$ monoclinic lattice distortion is a gradual process in FeGe while it is more drastic in iron-based superconductors. It originates from the different nature of the phase transition involved in the symmetry ascending, namely, the second-order-like phase transition in FeGe while it is first-order-like in iron-based superconductors.

We note that the CDW-related phonon modes disappear in the CDW phase and they cannot be explained by the phenomenological model of the $A_6^-$ mode and spin-canting interplay. One possible interpretation of the disappearing CDW-related phonon modes is their coupling to incommensurate spin fluctuations. From our inelastic neutron scattering experiments on FeGe, we find that the intensities of the low-energy incommensurate spin fluctuations do not follow the Bose population factor below $T_\text{CDW}$. Instead, they are enhanced dramatically below $T_\text{CDW}$~\cite{Chen_arxiv2023_neutron}. Although it is unclear why the CDW-induced phonons would disappear in the incommensurate magnetic ordered phase, it is possible that a coupling between the incommensurate static order and lattice can induce a secondary lattice instability below $T_\text{canting}$. As Raman scattering is a $Q=0$ probe, any small deviation from commensurate positions would have a large impact on Raman scattering results, but it would not dramatically influence the neutron and x-ray scattering results.

Compared with all the other kagome magnets, FeGe possesses magnetic order, CDW order, and a strong interaction between magnetic and CDW order~\cite{Teng_2022FeGe}. The CDW can also be tuned by a simple annealing process where the correlation length of CDW can change from 0 to 100\%~\cite{Chen_arxiv2023_STM,Wu_arxiv2023_Annealing,Shi_arxiv2023_Annealing}.  Beyond kagome systems, to the best of our knowledge, we are not aware of other CDW materials where the CDW can have such a strong coupling with magnetism and the CDW can also be tuned.
Thus, determining the symmetry of the crystal structure of FeGe is the very first step in sorting out the rich electronic and magnetic properties of this system.

The experimental observations in FeGe provide a rather rare type of spin-lattice coupling because the collinear A-type AFM structure is not expected to have an impact on the kagome lattice structure, particularly the in-plane crystal structure. We have shown that FeGe displays a series of structural phase transitions in the magnetic ordered phase, including threefold rotational symmetry breaking at $T_{\rm N}$, a CDW transition, negative thermal expansion, and a tendency of symmetry ascending at lower temperatures in the spin-canting phase, which are all directly related to the in-plane lattice distortion in the kagome structure. In particular, the lattice distortion is on the order of 10$^{-4}$; such a small lattice distortion is observed both by Raman and phase-sensitive Larmor neutron techniques.
In general, the energy scale of the crystal structural distortions is much larger than the magnetic exchange energy and spin-orbit coupling strength, as seen in cuprates, nickel oxides, and iron pnictides. FeGe seems to be a rare case where magnetic, lattice, and spin-orbit coupling energy scales are similar, resulting in their interaction and interplay.
This finding renders FeGe an interesting system where an extremely weak structural instability breaks the threefold symmetry, coexists, and competes with the CDW and magnetic orders.  
The unusual intertwined orders between spin, charge, and lattice degrees of freedom unveiled here may arise from the correlated electron effect of flattish electronic bands~\cite{Teng_arxiv2022_ARPES}.

\begin{acknowledgments}
We acknowledge useful discussions with Hengxin Tan and Binghai Yan.
The spectroscopic work conducted at Rutgers 
(S.-F.W. and G.B.) was supported by the National Science Foundation (NSF) Grants No.~DMR-2105001. 
The neutron scattering and single crystal synthesis work at Rice was supported by NSF-DMR-2100741 and by the Robert A. Welch Foundation under Grant No.~C-1839, respectively (P.D.).  
The theoretical work conducted at the University of Minnesota (J.S., E.R., and T.B.) was supported by the NSF~CAREER Grant No.~DMR-2046020.
X.K.T and M.Y. are partially supported by the Robert A. Welch Foundation Grant No.~C-2175, and the Gordon and Betty Moore Foundation’s EPiQS Initiative through grant No.~GBMF9470. 
The work at NICPB was supported by the European Research Council (ERC) under the European Union’s Horizon 2020 research and innovation program Grant Agreement No.~885413. A portion of this research used resources at the Spallation Neutron Source and High Flux Isotope Reactor, DOE Office of Science User Facilities operated by ORNL. The development of the Larmor diffraction technique was supported by the U.S. Department of Energy, Office of Science, Office of Basic Energy Sciences, Early Career Research Program Award (KC0402010), under Contract No. DE-AC05-00OR22725.

\end{acknowledgments}     
          
\cleardoublepage
             
\appendix 
                            
\section{Methods}\label{Methods}
                                                    
\textit{Single crystal preparation and characterization}\label{Crystal_preparation}
--                                                                                                             
Single crystals of FeGe were synthesized via the chemical vapor transport method described in Ref.~\cite{Teng_2022FeGe}, and the chemical compositions were determined by x-ray refinement~\cite{Teng_2022FeGe}. 
These samples were characterized by electric transport and magnetic susceptibility measurements.
The extracted collinear A-type antiferromagnetic phase-transition temperature $T_{N}$, CDW transition temperature ($T_\text{CDW}$), and spin-canting transition temperature ($T_\text{canting}$) for FeGe were about 400, 110, and 60\,K, respectively~\cite{Teng_2022FeGe}. 
The sharpness of the Raman modes and the low residual spectra background (Fig.~\ref{Fig4_T_dependence}) indicate the high quality of the single crystals.
                                                                                                                                
\textit{Raman scattering measurements}\label{Raman}
 --                                            
The as-grown FeGe sample with the (0~0~1) surface was positioned in
a continuous-helium-flow optical cryostat.  The Raman data shown in the main text are obtained from this sample.
A polished (0~0~1) surface and an as-grown (1~0~0) surface of the FeGe crystals were also studied, and they showed consistent results.
The Raman measurements
were mainly performed using the Kr$^+$ laser line at 647.1\,nm (1.92\,eV) in
a quasibackscattering geometry along the crystallographic $c$ axis. 
The excitation laser beam was focused into a $50\times100$ $\mu$m$^2$
spot on the $ab$ surface, with an incident power around 17\,mW. The
scattered light was collected and analyzed by a triple-stage Raman
spectrometer, and recorded using a liquid-nitrogen-cooled
charge-coupled detector. 
Linear and circular polarizations were used in this study to decompose the Raman data into different irreducible representations.
The instrumental resolution was maintained better than 1.5\,\cm-1.
All linewidth data presented were corrected for the instrumental resolution. 
The temperatures shown in this paper were corrected for laser heating (Appendix~\ref{laser_heating_determination}).

All spectra shown were corrected for the spectral response of the spectrometer and charge-coupled detector to obtain the Raman intensity $I
_{\mu v}$, which is related to the Raman response $\chi''(\omega,T)$: $I_{\mu v}(\omega, T)=[1+n(\omega, T)] \chi_{\mu \nu}^{\prime \prime}(\omega, T)$. Here, $\mu (v)$ denotes the polarization of the 
incident (scattered) photon, $\omega$ is the energy, $T$ is the temperature, and $n(\omega, T)$ is the Bose factor.
                                                                                                                                                                                                        
The Raman spectra were recorded from the $ab$ (0~0~1) surface for scattering geometries denoted as $\mu v = XX, XY, RR, RL$, which is short for $Z(\mu v)\bar{Z}$ in Porto’s notation, where $X$ and $Y$ denote linear polarization parallel and perpendicular to the crystallographic $a$ or $b$ axis, respectively, and $R=X+iY$ and $L=X-iY$ denote the right- and left-circular polarizations, respectively. The $Z$ direction corresponds to the $c$-axis direction perpendicular to the (0~0~1) plane. 
                                 
\textit{Neutron Larmor diffraction measurement}
--
Neutron Larmor diffraction measurements were performed on the HB-1 polarized triple axis spectrometer of the High Flux Isotope Reactor (HFIR) at Oak Ridge National Laboratory (ORNL), USA. We used a single crystal (about 100\,mg) mounted inside a closed-cycle refrigerator with an operating temperature range between 20 and 460\,K. It is a bulk measurement and probes the entire sample. The momentum transfer ${\bf Q}$ in 3D reciprocal space in \AA$^{-1}$ was defined as 
${\bf Q}=H{\bf a}^\ast+K{\bf b}^\ast+L{\bf c}^\ast$, where $H$, $K$, and $L$ are Miller indices with ${\bf a}=a{\bf \hat{x}}$, ${\bf b}=b(\cos120 {\bf \hat{x}}+\sin120 {\bf \hat{y}})$, and ${\bf c}=c {\bf \hat{z}}$ ($a\approx b\approx 4.99$ \AA, $c\approx 4.05$ \AA\ at room temperature).  Although this definition of an ideal kagome lattice structure, strictly speaking, is only valid above $T_{\rm N}$, we use this notation throughout the paper as the in-plane lattice distortion is small enough that it does not affect the discussion.

\textit{Density functional theory calculations}\label{DFT}
 --  
DFT calculations were performed within the Perdew-Burke-Ernzerhof-type generalized gradient approximation~\cite{Perdew1996PhysRevLett}, which is implemented in the Vienna $Ab$ initio Simulation Package (VASP)~\cite{Kresse1996_PhysRevB,KRESSE199615} using the experimental crystal structure. The projected augmented wave potentials with nine valence-electrons for the Fe atom and five valence electrons for Ge were employed. The cutoff energy for the
plane-wave basis set was 300\,eV. The zero-damping DFT-D3 van der Waals correction was employed throughout the calculations. The phonon dispersion was calculated by using the finite displacement method as implemented
in the phonopy code~\cite{TOGO20151}. The on-site Coulomb interaction $U$ was set to 2\,eV for Fig.~\ref{Fig_Raman_DFT}(b). For the phonon calculation in the nonmagnetic phase, we did not introduce a magnetic order.  

Group theory predictions were performed using the tool provided in the Isotropy Software Suite and the Bilbao Crystallographic Server \cite{Bilbao_1, Bilbao_4, Hatch2003}. The information for the irreducible representations of point groups and space groups follow the notations of Cracknell, Davies, Miller, and Love~\cite{cracknell1979general}.

\section{Laser heating determination} \label{laser_heating_determination}                                                                                                                             
The laser heating rate, a measure of the temperature increase per unit laser power (K/mW) in the focused laser spot, in the Raman experiments was determined by monitoring 
the appearance of new phonon modes induced by the CDW order during the cooling process with a constant laser power of 17\,mW.
             
At the cryostat temperature 95\,K, we barely detect any new phonon modes, indicating the laser spot temperature is above $T_\text{CDW}$=110\,K.
When cooling the sample to 90\,K, we start to detect several weak new-phonon signals both in the $RR$ and $RL$ scattering geometries, indicating the laser spot temperature is slightly below 110\,K. When cooling the sample to 85\,K, the intensity of these new modes develops significantly, indicating the laser spot temperature is well below 110\,K. 
Thus, the heating coefficient can be determined via $90\,\text{K}+17\,\text{mW}*\text{k} \approx 110$\,K. In this way, we have deduced the heating coefficient $k \approx 1.2\pm0.1$\,K/mW.

We note that the heating coefficient $k$ is not a constant with cooling. The $T$ dependence of the heating coefficient $k$ can be estimated by solving the heat transfer equation. The thermal conductivity, incident laser power $P$, and temperature of interest inside the laser spot $T_\text{spot}$ are connected by the integral equation~\cite{MAKSIMOV1992407,Ye_2021_PhysRevB}
\begin{equation}
\int_{T_0}^{T_{\text {spot }}}\kappa(T) d T=\frac{P \cdot d^*}{S}=\mathrm{constant}
\end{equation}
where $T_0$ is the cold helium-gas temperature where the sample is located in the cryostat, $S$ is the area of the laser spot, and $d^*$ is an effective thickness. The constant $P \cdot d^*/S$ can be determined by a single measurement of a distinctive temperature in the laser spot, as we did at $T_0=90$\,K, where $T_\text{spot}=110$\,K.


Since the thermal conductivity data of FeGe are not available in the literature, we estimated it from the measured in-plane electric conductivity data $\sigma (T)$, which is connected by the in-plane resistivity as $1/\rho(T)$~\cite{Teng_2022FeGe}.
Based on the Wiedemann-Franz law for a simple metal, the thermal conductivity $\kappa(T)$ can by approximated by $\sigma(T)*T*L_0$, where $L_0$ is the Lorenz number $2.44 \times 10^{-5}$ mW$\Omega$K$^{-2}$. We note that the Wiedemann-Franz law is generally valid for high temperatures and for low (i.e., a few Kelvins) temperatures but may not hold at intermediate temperatures~\cite{rosenberg1988solid}. Nevertheless, we roughly estimate the $T$-dependent heating coefficient assuming that $L_0=\kappa(T)/(\sigma(T)*T)$ is a constant for FeGe. The estimated $T$-dependent heating coefficients are shown in  Table~\ref{heating}.

For $T_0$=10\,K and 20\,K, the heating coefficient changes substantially (less than 1.6 times) compared with $T_0=$90\,K. For a wide temperature range from $T_0=20$\,K to 300\,K, the heating coefficient does not vary much compared with $T_0=$90\,K.

\begin{table}[b]
\caption{\label{heating}  
\textcolor{black}{  
Cryostat temperature $T_0$, laser spot temperature $T_\text{spot}$, laser heating $\Delta T$, the heating coefficient $k$ for FeGe at different temperatures with a constant laser power of $P=17$\,mW.}
}
\begin{ruledtabular}
\begin{tabular}{cccc}
 $T_0$(K) &$T_\text{spot}$(K) & $\Delta T$(K)    &  $k$(K/mW)   \\
\hline
10&42.5&32.5&1.91\\
20&46.6&26.6&1.56\\
40&60.8&20.8&1.22\\
90&110&20&1.18\\
150&172&22&1.29\\
200&223&23&1.35\\
300&323&23&1.35\\
\end{tabular}
\end{ruledtabular}
\end{table}  
                           
%
%
        

              
\section{Simulation of the lattice Bragg peaks in the momentum space}\label{Bragg_peaks} 
                      
In this appendix, we present the simulation of the three lattice Bragg peaks (2,~0,~0), (0,~2,~0), and (2,~-2,~0) in the $k_z=0$ momentum space for FeGe. These three lattice Bragg peaks correspond to three lattice spacings in real space as shown in Fig.~\ref{Fig2_AFM}(g). The simulation is performed using the ISODISTORT program from ISOTROPY Software Suite~\cite{Stokes_web,Campbell_wf5017}.
     
\begin{figure*}[!t] 
\begin{center}
\includegraphics[width=2\columnwidth]{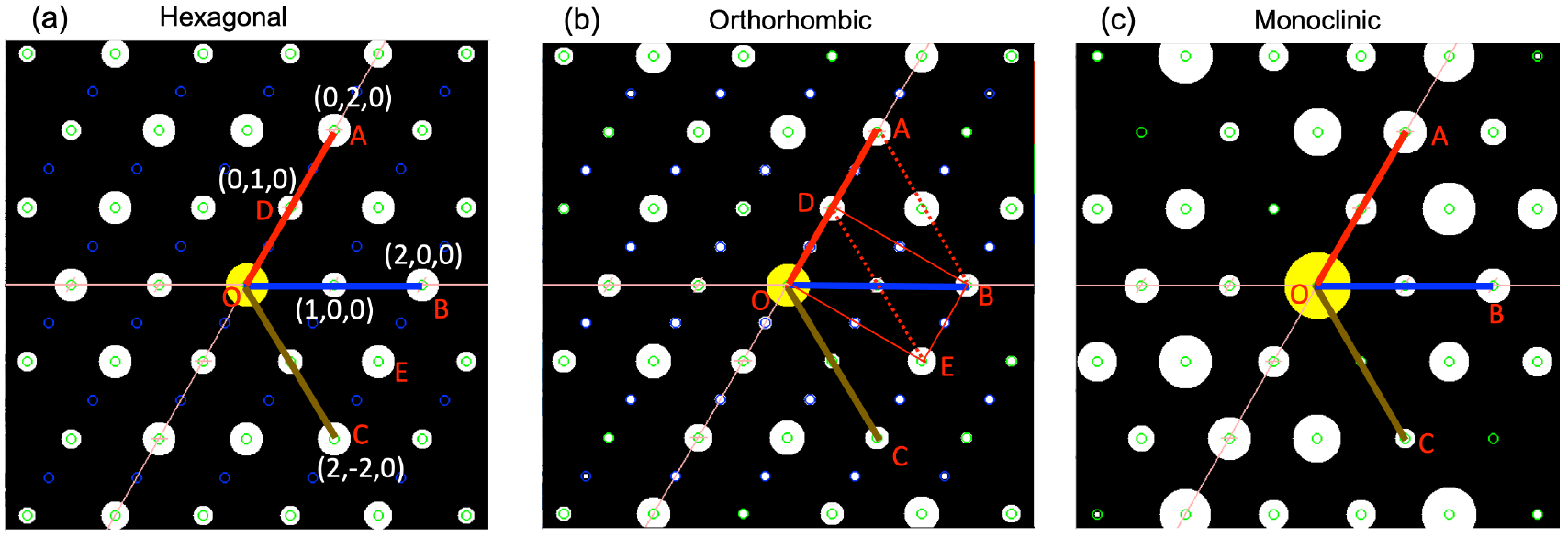}
\end{center} 
\caption{\label{Bragg_points} 
Lattice Bragg peaks simulation in the momentum space for the hexgonal phase (a), orthorhombic phase (b), and monoclinic phase (c) in the $k_z=0$ plane using ISODISTORT program from ISOTROPY Software Suite~\cite{Stokes_web,Campbell_wf5017}. The Lattice Bragg peaks (0,~2,~0), (2,~0,~0), (2,~-2,~0), (0,~1,~0), (2,~-1,~0), and (0,~0,~0) are represented by A, B, C, D, E, and O, respectively. The distance of $OA$, $OB$, and $OC$ are marked by solid red, blue, and brown lines following the same color scheme in Fig.~\ref{Fig2_AFM}(d), respectively. 
}
\end{figure*}

For the nonmagnetic hexagonal phase shown in Fig.~\ref{Bragg_points}(a), threefold symmetry is preserved; thus, $OA=OB=OC$. For both the orthorhombic and monoclinic phases of FeGe, they can be driven by either $A_6^-$ or $\Gamma_6^-$ lattice instabilities (Appendix~\ref{Group_subgroup_analysis}). Note that $A_6^-$ and $\Gamma_6^-$ lattice instabilities give rise to a similar unit-cell shape. The difference is that $A_6^-$ leads to unit-cell doubling along the $c$ axis while $\Gamma_6^-$ does not. Since we focus on the  $k_z=0$ plane, the $A_6^-$ and $\Gamma_6^-$ lattice instabilities contribute to similar Bragg peak structures in momentum space.
                        
In Fig.~\ref{Bragg_points}(b), we show the simulation of the orthorhombic phase in momentum space. Because of the in-plane unit-cell doubling [Fig.~\ref{Fig1_intro}(c)], new Bragg peaks appear in either the $OA$, $OB$, or $OC$ directions corresponding to different domain orientations. For illustration purposes, we show that new Bragg peaks appear in the $OA$ direction, e.g., at half of $OD$ and equivalent positions. The rectangular shape of the $OEBD$ restricts the diagonal of the rectangle to be equal, namely, $DE=OB$. Since $DE=OC$, we obtain $OB=OC$. Because of the threefold symmetry breaking, $OB$ and $OC$ deviate from $OA$. As a consequence, we obtain the relation $OB=OC \neq OA$.

For the monoclinic phase, the symmetry is lower than the orthorhombic phase. The restriction that $OB=OC$ is removed. Thus, we obtain $OB \neq OC \neq OA$.
                    
\section{Structure factor analysis for the (2~0~0) and its equivalent Bragg peaks}\label{Structure_factor_analysis}

In the undistorted hexagonal phase, the Fe and Ge atoms all occupy the high-symmetry positions (Wyckoff position: $3f$ for Fe, $1a$ for Ge$^1$, and $2d$ for Ge$^2$). The neutron scattering structure factor for a general nuclear Bragg peak $\mathbf{Q_{\textit{HKL}}}=(H,K,L)$ is given by
\begin{align*}
    &F_{\text{hex}}(\textit{H,K,L}) \\
   &=b_{\text{Fe}}\Sigma_j^3e^{i\mathbf{Q_{\textit{HKL}}\cdot\mathbf{R(Fe^j)}}}+b_{\text{Ge}}\Sigma_j^3e^{i\mathbf{Q_{\textit{HKL}}\cdot\mathbf{R(Ge^j)}}}\\
    &=b_{\text{Fe}}\left[e^{i2\pi(H/2)}+e^{i2\pi(K/2)}+e^{i2\pi(H/2+K/2)}\right]\\
    &+b_{\text{Ge}}\left[e^{i2\pi(0)}+e^{i2\pi(H/3+2K/3+L/2)}+e^{i2\pi(2H/3+K/3+L/2)}\right]
\end{align*}
Here, $b_{\text{Fe}}$ and $b_{\text{Ge}}$ are the neutron scattering lengths for the Fe and Ge nuclei, respectively, and $\mathbf{R}$(Fe$^j$) and $\mathbf{R}$(Ge$^j$) are the fractional coordinates for Fe and Ge, respectively. Note that $j$ is the index for the three Fe and Ge atoms in one unit cell. For $(2,0,0)$ and its equivalent Bragg peaks, the above formula results in the following:
\begin{equation*}
    F_{\text{hex}}(2,0,0)=3\cdot b_{\text{Fe}}+0\cdot b_{\text{Ge}}
\end{equation*}
It is clear that in the undistorted hexagonal phase, the $(2,0,0)$ and its equivalent Bragg peaks only have a Fe contribution, because the three Ge atoms in the unit cell add up destructively at this particular reciprocal space position while the three Fe atoms add up constructively.

The above equations for the structure factor still hold in the case of the slightly distorted lattice. The difference is that in the distorted lattice, the site symmetry for both Fe and Ge atoms will be lower compared with the hexagonal phase. Thus, the in-plane coordinates of Fe and Ge are no longer protected by the point-group symmetry of the hexagonal phase and can deviate from the high-symmetry fractional numbers of $\frac12, \frac13$, and $\frac23$. In this case, the Ge contribution to the $(2,0,0)$, $(0, 2, 0)$, and $(2, -2, 0)$ Bragg peaks will not exactly cancel out. Since the lattice distortion obtained from the Larmor diffraction measurement is of the order of $10^{-4}$, the Ge component at those Bragg positions should also be approximately $10^{-4}$ compared to the Fe component. Because the lattice spacing difference obtained in Larmor data is based on measurements of $(2,0,0)$, $(0, 2, 0)$, and $(2, -2, 0)$ Bragg peaks, these results are mainly affected by the Fe sublattice in the kagome plane.

\section{Group-theoretical analysis of the Raman-active modes of the nonmagnetic FeGe}\label{GroupTheory}                            
                                                                                                                                                     
In this appendix, we discuss the group-theoretical analysis of the phonon modes in the high-temperature nonmagnetic phase ($T>T_\text{N}$).                                                                                                                                                                                                                                                                                                                                                                                         
The high-temperature nonmagnetic FeGe belongs to the hexagonal structure with space group $P6/mmm$ (No.~191) (point group:~$D_{6h}$). 
The Fe, Ge$^1$ (in the kagome layer), and Ge$^2$ atoms (in the honeycomb layer) have Wyckoff positions $3f$, $1a$, and $2d$, respectively.                                                             
From the group theoretical considerations~\cite{Bilbao_1}, $\Gamma$-point phonon modes of the hexagonal nonmagnetic FeGe can be expressed as $\Gamma_\text{total}$ = 3$A_{2u}$ $\oplus$ $B_{2g}$ $\oplus$ $B_{1u}$ $\oplus$ $B_{2u}$ $\oplus$ $E_{2u}$ $\oplus$ $E_{2g}$ $\oplus$ 4$E_{1u}$. Raman active modes are $\Gamma_{\text{Raman}}$= $E_{2g}$, IR active modes are $\Gamma_{\text{IR}}$=2$A_{2u}$ $\oplus$ 3$E_{1u}$, the acoustic mode is $\Gamma_{\text{acoustic}}$ =$A_{2u}$ $\oplus$ $E_{1u}$, and the silent modes are $\Gamma_{\text{silent}}$ = $B_{2g}$ $\oplus$ $B_{1u}$ $\oplus$ $B_{2u}$ $\oplus$ $E_{2u}$.
Note that 
(1) the Raman-active $E_{2g}$ mode is related to the in-plane lattice vibrations of the Ge$^2$ atoms of the honeycomb layer;
(2) $E_{2g}$ and $A_{1g}$ modes can be accessed from the $ab$-plane measurement while the $E_{1g}$ mode can only be accessed from the side surface measurement; (3) the $A_{1g}$ and $E_{1g}$ modes are not symmetry allowed in the high-temperature nonmagnetic phase; and (4) there are no Raman-active modes for Fe and Ge$^1$ atoms in the kagome layer of the high-temperature nonmagnetic phase.

\section{Robustness of the $A_6^-$ lattice instability}\label{DFT_robustness}

\begin{table}[b]
\caption{\label{A_point_instabilities} Three lowest phonon modes at the $A$ point based on the DFT phonon calculations at different ISMEAR and SIGMA parameters.}
\begin{ruledtabular}
\begin{tabular}{cccc|ccc}
	Smearing & Smearing & $\mathrm{U}$ & Magnet-& \multicolumn{3}{c}{Softest $A$ } \\
Method& Width $(\mathrm{eV})$ & $(\mathrm{eV})$ & ism?& \multicolumn{3}{c}{point modes ($\mathrm{c m}^{-1}$)} \\
 \hline
 Gaussian & 0.05 & 0 & - & $-172$ & 105 & 210 \\
 Gaussian & 0.20 & 0 & - & $-71$ & 118 & 216 \\
 Fermi & 0.05 & 0 & - & $-193$ & 104 & 210 \\
 Fermi & 0.10 & 0 & - & $-165$ & 105 & 210 \\
 Fermi & 0.20 & 0 & - & $-123$ & 109 & 212 \\
 Fermi & 0.20 & 1 & - & $-128$ & 108 & 212 \\
 Fermi & 0.20 & 2 & - & $-131$ & 107 & 213 \\
 Fermi & 0.20 & 3 & - & $-104$ & 110 & 218 \\
 Fermi & 0.20 & 0 & AFM & 78 & 168 & 248 \\
 Fermi & 0.20 & 1 & AFM & 82 & 177 & 262 \\
 Fermi & 0.20 & 2 & AFM & 87 & 184 & 277 \\
 Fermi & 0.20 & 3 & AFM & 90 & 191 & 291 \\
\end{tabular}
\end{ruledtabular}
\end{table} 
     
In this appendix, we show that the $A_6^-$ lattice instability is robust for a reasonable range of parameters in the DFT calculations.
                                                                     
Two input parameters, which are not well determined but can affect the outcome of this type of first-principles calculation, are the Hubbard +U and the Fermi surface smearing parameter $\sigma$. The +U correction helps to capture the on-site repulsion between the $d$ electrons for the transition metal, and the $\sigma$ parameter determines how the partial occupancies near the Fermi level are treated. In systems where the lattice instabilities are intertwined and possibly driven by the Fermi surface effects, $\sigma$ can lead to very large changes for the unstable mode frequencies. As a result, different values of $\sigma$ should be examined to ensure the robustness of the reported unstable modes. 
         
In Table~\ref{A_point_instabilities}, we show the lowest three phonon frequencies at the $A$ point with different $\sigma$ values and smearing types (the ``ISMEAR'' tag in VASP, which determines Gaussian vs Fermi smearing). In the absence of a magnetic order, we always find an $A_6^-$ instability. This calculation is performed in a $1\times 1\times 2$ supercell using the frozen-phonon approach in order to avoid the possible errors in the procedure used to obtain phonon dispersions. 
When the calculation is performed without spin polarization, there is a single $A$-point mode that is sensitive to the smearing width but is consistently unstable. When the magnetically ordered (AFM) phase is considered, this mode becomes stable. The absence of an $A$-point instability in the AFM phase is consistent with the DFT calculations shown in Refs.~\cite{Shao_2022FeGeDFT,Miao_arxiv2022_xray,Ma_arxiv2023_DFT,Wang_arxiv2023_DFT}, and it suggests that the instability at the $A$ point is very sensitive to the magnetic order.

One drawback for a DFT phonon calculation without taking the magnetic order into account is that fictitious instabilities may emerge throughout the Brillouin zone.
While this is not the case for the phonon dispersions presented in Fig.~\ref{Fig_Raman_DFT}(b), we performed frozen-phonon calculations in a $2\times 2\times 2$ supercell to confirm this observation and extract the representations of the instabilities. This supercell is commensurate with the $\Gamma$, $M$, $L$, and $A$ points of the Brillouin zone. In Table~\ref{Unstable_phonon_modes}, we list the softest phonon frequencies on these points. There are no other instabilities at high-symmetry points other than those in Fig.~\ref{Fig_Raman_DFT}(b).
     
\begin{table}[b]
	\caption{\label{Unstable_phonon_modes} Unstable phonon modes and their irreducible representations for the high-symmetry points in $k$ space based on the DFT phonon calculation in a $2\times 2\times 2$ supercell. (No magnetic order was imposed, and $U = 2$\,eV was used.)}
\begin{ruledtabular}
\begin{tabular}{ccc}
	Frequency (\cm-1)& Degeneracy &Representation \\
 \hline
$98 $  &3 & $L_2^-$ \\
$71 $  &2 & $\Gamma_5^-$ \\
$63 $  &1 & $A_3^-$ \\
$35 $  &3 & $L_3^-$ \\
$21 $  &3 & $M_ 2^+$ \\
$0 $   &2 & $\Gamma_6^-$ \\
$0 $   &1 & $\Gamma_2^-$ \\
$38i $  &1 & $\Gamma_3^-$ \\
$45i $  &3 & $M_3^-$ \\
$113i $ &2 & $\Gamma_6^-$ \\
$132i $ &2 & $A_6^-$ \\
\end{tabular}
\end{ruledtabular}
\end{table}

\section{Group-subgroup analysis}\label{Group_subgroup_analysis} 

In Tables~\ref{table:subgroup1} and \ref{table:subgroup2}, we show the low-symmetry space groups that can be obtained from the parent P6/mmm by the symmetry-breaking  $\Gamma_6^-$ and $A_6^-$ irreducible representations (irreps). 

The $A_6^-$ instability can drive a transition to orthorhombic space groups Cmcm (\#63, point group $D_{2h}$) or  monoclinic P2$_1$/m (\#11, point group $C_{2h}$).
The $\Gamma_6^-$ instability, if it freezes in and drives the transition to a lower-symmetry phase, leads to either the orthorhombic space group Amm2 (point group $C_{2v}$) or the monoclinic space group Pm (point group $C_s$). The conclusion is the same: that both $A_6^-$ and $\Gamma_6^-$ lead to either the orthorhombic or the monoclinic phase, breaking the threefold symmetry at $T_N$. The difference is that $A_6^-$ leads to a doubling of the unit cell along the $c$-axis direction while $\Gamma_6^-$ does not. However, all of the low-symmetry structures have a large number of Raman-active modes, most of which are too weak to give rise to detectable mode intensity. Thus, a $\Gamma_6^-$ lattice instability does not fundamentally change our conclusions, except that 
the interlayer $c$-axis dimerlike singlet cannot be formed due to the $\Gamma_6^-$ mode.

\begin{table}[t]
\begin{tabular}{|c|c|c|c|}
	\hline
	Direction	& Space group	& Fe sites & Ge sites\\
	\hline
	$A_6^-$(a,0)	& Cmcm (\# 63)		& 4c, 8g & 4c, 8f\\
	$A_6^-$(0,a)	& Cmcm (\# 63)		& 4c, 8g & 4c, 8e\\
	$A_6^-$(a,b)	& P2$_1$/m (\#11)	& 2e, 2e, 2e & 2e, 4f\\
	\hline
\end{tabular}
	\caption{Low-symmetry space groups that can be obtained from the parent P6/mmm by the symmetry-breaking $A_6^-$ irrep. }
	\label{table:subgroup1}
\end{table}

\begin{table}[t]
\begin{tabular}{|c|c|c|c|}
        \hline
        Direction       & Space group   & Fe sites & Ge sites\\
        \hline
        $\Gamma_6^-$(0,a)		    & Amm2 (\# 38)          & 2a, 4d & 2a, 2b, 2b\\
	$\Gamma_6^-$(a,$\sqrt{3}$a)	    & Amm2 (\# 38)          & 2a, 4d & 2a, 4e\\
        $\Gamma_6^-$(a,b) 		    & Pm (\#6)       	    & 1a, 1a, 1a & 1a, 1b, 1b\\
        \hline
\end{tabular}
        \caption{Low-symmetry space groups that can be obtained from the parent P6/mmm by the symmetry-breaking $\Gamma_6^-$ irrep. }
	        \label{table:subgroup2}
\end{table}

\section{$A_6^-$ displacement}\label{A6appendix} 
\begin{figure}[!t] 
\begin{center}
\includegraphics[width=0.8\columnwidth]{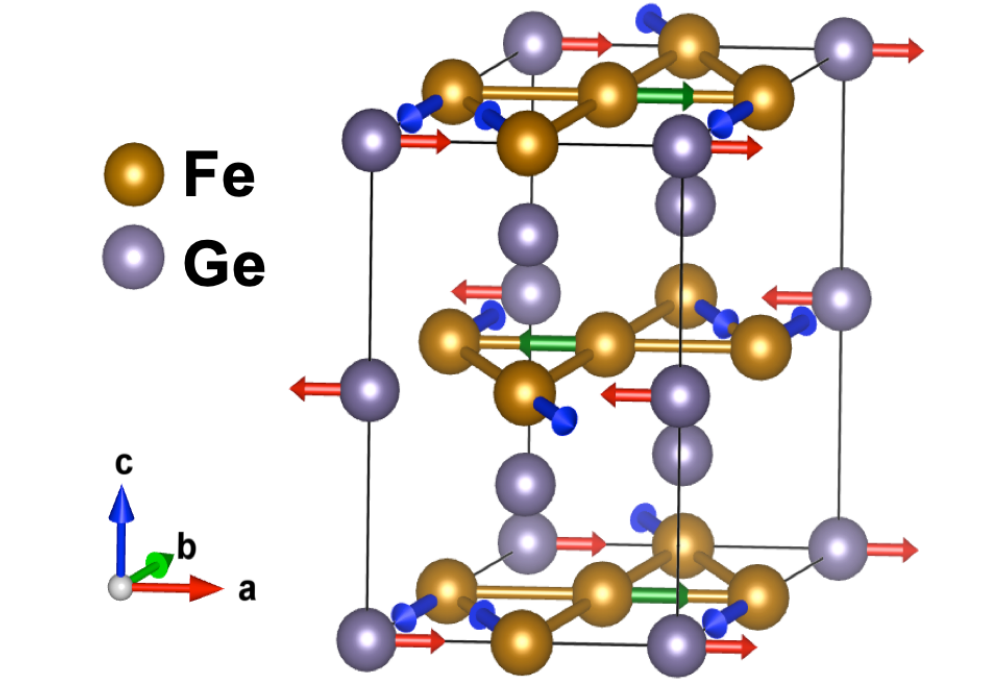}
\end{center}
\caption{\label{A6-} 
The $A_6^-$ displacement pattern of the nonmagnetic phase based on the DFT phonon calculation.}
\end{figure}

In this appendix, we present the $A_6^-$ displacement pattern for the nonmagnetic hexagonal phase shown in Fig.~\ref{A6-} based on the DFT phonon calculations for the nonmagnetic phase.

\section{Raman tensor analysis}\label{Raman_tensor_analysis}  
                   
The Raman tensor $R_{\mu}$ for an irreducible representation ($\mu$) of a point group is a $2\times 2$ matrix.
With the unit vectors for the polarization of the incident light ($\hat{e}_i$) and scattering light ($\hat{e}_s$), the phononic Raman response is described in the following way:
\begin{equation}
\label{eq_chi}
\chi''_{\hat{e}_i \hat{e}_s}=\sum_\mu|\hat{e}_i R_\mu \hat{e}_s |^2.
\end{equation}

%

\subsection{$D_{6h}$ Raman tensor}\label{D6h}  
The Raman tensors $R_{\mu}$ ($\mu=A_{1g}, A_{2g}, E_{1g}, E_{2g}$) for the irreducible representations ($\mu$) of point group $D_{6h}$ have the following forms:
\begin{displaymath}   
\left(\begin{array}{ccc}
a & 0  &0\\
0 &a &0\\
0 & 0 &b
\end{array}\right),\\
\left(\begin{array}{ccc}
0 & c  &0\\
-c &0 &0\\
0 & 0 &0
\end{array}\right),
\left(\begin{array}{ccc}
0 & 0  &0\\
0 &0 &d\\
0 & e &0
\end{array}\right)
\left(\begin{array}{ccc}
0 & 0  &-d\\
0 &0 &0\\
-e & 0 &0
\end{array}\right),\\
\end{displaymath}  
\begin{displaymath}   
\left(\begin{array}{ccc}
0 & f  &0\\
f &0 &0\\
0 & 0 &0
\end{array}\right)
\left(\begin{array}{ccc}
f & 0  &0\\
0 &-f &0\\
0 & 0 &0
\end{array}\right).
\end{displaymath}

We choose $\hat{e}_i$ and $\hat{e}_s$ to be $X$, $Y$, $R$, and $L$, where $X=(1~0~0)$, $Y=(0~1~0)$, $R=1/\sqrt{2} ~(1~i~0)$, and $L=1/\sqrt{2}~(1~-i~0)$. 

Based on Eq.~(\ref{eq_chi}), we obtain: 
\begin{equation}
\label{scattering_geometryD6h}
\begin{split}     
\chi''^{D_{6h}}_{XX}&=a^2+f^2,\\
\chi''^{D_{6h}}_{XY}&=c^2+f^2,\\
\chi''^{D_{6h}}_{RR}&=a^2+c^2,\\
\chi''^{D_{6h}}_{RL}&=2f^2.\\
\end{split}
\end{equation}
Thus, the Raman selection rules for the $D_{6h}$ point group indicate that the $XX$, $XY$, $RR$, and $RL$ polarization geometries probe the $A_{1g} + E_{2g}$, $A_{2g} + E_{2g}$, $A_{1g} + A_{2g}$, and 2$E_{2g}$ symmetry excitations, respectively~(Table~\ref{SymmetryAnalysis}). 
                                                                                                                   
The sum rule that 
$\chi''^{D_{6h}}_{XX}$ + $\chi''^{D_{6h}}_{XY}$ = $\chi''^{D_{6h}}_{RR}$ + $\chi''^{D_{6h}}_{RL}$ = $a^2+c^2+2f^2$
sets a constraint for the Raman response in different scattering geometries, 
thus providing a unique way to check the data consistency.
                                                                                                                                                 
From Eq.~(\ref{scattering_geometryD6h}), we can calculate the square of the Raman tensor element:
\begin{equation}
\label{decompossiitonD6hEQ}
\begin{split}     
a^2&=\chi''^{D_{6h}}_{XX}-\chi''^{D_{6h}}_{RL}/2,\\
c^2&=\chi''^{D_{6h}}_{XY}-\chi''^{D_{6h}}_{RL}/2,\\
f^2&=\chi''^{D_{6h}}_{RL}/2.\\
\end{split}
\end{equation}
Therefore, the algebra in Eq.~(\ref{decompossiitonD6hEQ}) can be used to decompose the measured Raman signal into three separate irreducible representations ($A_{1g}$, $A_{2g}$, $E_{2g}$) of the point group $D_{6h}$~(Table~\ref{decompositionD6h}).  
                                                                                                                                                                           
This decomposition algebra is a characteristic property of a lattice system with trigonal or hexagonal symmetry, where the threefold rotational symmetry is preserved. Whether sixfold rotational symmetry is preserved or not depends on the system.
                                                                      
\begin{table}[t]
\caption{\label{SymmetryAnalysis} Relationship between the scattering geometries and the symmetry channels. Here, $A_{1g}$,  $A_{2g}$, and $E_{2g}$ are the irreducible representations of the $D_{6h}$ point group.
}
\begin{ruledtabular}
\begin{tabular}{cc}
Scattering geometry&Symmetry channel\\
\hline
$XX$&$A_{1g}+E_{2g}$\\
$XY$&$A_{2g}+E_{2g}$\\
$RR$&$A_{1g}+A_{2g}$\\
$RL$&$2E_{2g}$\\
\end{tabular}
\end{ruledtabular}
\end{table}
     
\begin{table}[t]
\caption{\label{decompositionD6h} Algebra used to decompose the Raman data into three irreducible representations of the point group $D_{6h}$.}
\begin{ruledtabular}
\begin{tabular}{cc}
Symmetry channel&Expression\\
\hline
$A_{1g}$&$\chi''_{XX}-\chi''_{RL}/2$\\
$A_{2g}$&$\chi''_{XY}-\chi''_{RL}/2$\\
$E_{2g}$&$\chi''_{RL}/2$\\
\end{tabular}
\end{ruledtabular}
\end{table}

\begin{table}[t]
\caption{\label{SymmetryAnalysiSoD2h} Relationship between the scattering geometries and the symmetry channels for the $C_{2h}$ Raman tensor. Here, $A_{g}$ and $B_{g}$ are the irreducible representations of the $C_{2h}$ point group. }
\begin{ruledtabular}
\begin{tabular}{cc}
Scattering geometry&Symmetry channel\\
\hline
$XX$&$A_{g}$\\
$YY$&$A_{g}$\\
$XY$&$B_{g}$\\
$RR$&mainly $A_{g}$\\
$RL$&mainly $B_{g}$\\
\end{tabular}
\end{ruledtabular}
\end{table}

\subsection{$C_{2h}$ Raman tensor}\label{C2h2}  
                 
The Raman tensors $R_{\mu}$ for the irreducible representation of point group $C_{2h}$ ($\mu=A_{g}, B_{g}$) have the following form:
\begin{displaymath}   
\left(\begin{array}{ccc}
p & t  &0\\
s &q &0\\
0 & 0 &r
\end{array}\right),\\
\left(\begin{array}{ccc}
0 & 0  &w\\
0 &0 &r\\
u & v &0
\end{array}\right).
\end{displaymath}  
                                                                                                             
                                    
For the $A_6^-$-driven AFM monoclinic phase, the $C_2$ axis is perpendicular to the threefold axis of the nonmagnetic phase.
In this case, we choose $\hat{e}_i$ and $\hat{e}_s$ to be $X$, $Y$, $R$, and $L$, where $X=(1~0~0)$, $Y=(0~0~1)$, $R=1/\sqrt{2} ~(1~0~i)$, and $L=1/\sqrt{2}~(1~0~-i)$. Note that the choice of $\hat{e}_i$ and $\hat{e}_s$ is with respect to the monoclinic Raman tensor where the $C_2$ axis lies in the $ab$ plane.
Following Eq.~(\ref{eq_chi}), we obtain
        
\begin{equation}
\label{scattering_geometryC2h}
\begin{split}     
\chi''^{C_{2h}}_{XX}&=p^2,\\
\chi''^{C_{2h}}_{YY}&=r^2,\\
\chi''^{C_{2h}}_{XY}&=w^2,\\
\chi''^{C_{2h}}_{RR}&=1/4 ((p + r)^2 + (u - w)^2),\\
\chi''^{C_{2h}}_{RL}&=1/4 ((p - r)^2 + (u + w)^2).\\
\end{split}
\end{equation}
The Raman selection rules for the $C_{2h}$ point group in a single-domain sample indicate that $XX$, $YY$, and $XY$ polarization geometries probe the $A_{g}$, $A_{g}$, and $B_{g}$ symmetry excitations, respectively~(Table~\ref{SymmetryAnalysiSoD2h}).

Since the monoclinic lattice distortion is about $3\times 10^{-4}$ at room temperature in the AFM phase [Fig.~\ref{Fig2_AFM}(g)], the anisotropy of the Raman tensor elements would be tiny.
We thus obtain $p \sim r$ and $u \sim w$. According to Eq.~(\ref{scattering_geometryC2h}), $RR$ mainly probes the $A_g$ symmetry excitations while $RL$ mainly probes the $B_{g}$ symmetry excitations. 

\begin{figure}[!t] 
\begin{center}
\includegraphics[width=\columnwidth]{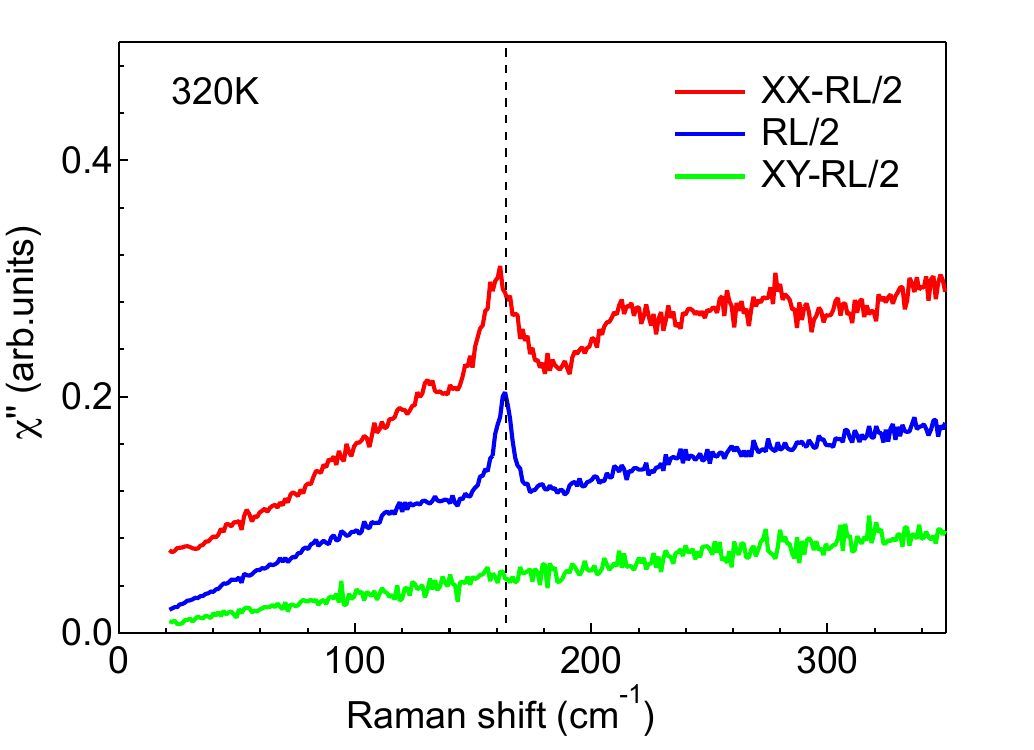}
\end{center}
\caption{\label{decomposition_320K} 
Symmetry decomposition into separate channels using the decompositions rules shown in Table~\ref{decompositionD6h} for FeGe at 320\,K.
}
\end{figure}

In the case of $p \sim r$ and $u \sim w$, the decomposition rule recovers to the $D_{6h}$ case shown in Table~\ref{decompositionD6h}.
The two modes in between 160 and 162\cm-1 can be separated using the decomposition rules shown in Table~\ref{decompositionD6h}.
As we show in Fig.~\ref{decomposition_320K}, a broader peak at around 160\,\cm-1 in the $XX-RL/2$ channel and a sharper mode at 162\,\cm-1 in the $RL/2$ channel can be clearly seen, while there is barely any phonon intensity in the $XY-RL/2$ channel.
The application of the decomposition rule according to the point group $D_{6h}$ for FeGe additionally confirms that  the AFM monoclinic phase only slightly deviates from the hexagonal lattice.

\section{Phonon peak positions in the CDW phase}\label{Phonon_peak_position_CDW_phase}
In this appendix, we present a summary of the phonon peak positions in the CDW phase at 90\,K in Table~\ref{phonon_modes222}. 
\begin{table}[b]
\caption{\label{phonon_modes222} Experimentally observed phonon frequencies at the Brillouin zone center for FeGe in the CDW phase at 90\,K. All the units are in \cm-1.}
\begin{ruledtabular}
\begin{tabular}{cc}
Scattering geometry &Frequency \\
\hline
&100\\
&127\\  
$RR$&174\\  
&203\\  
&249\\   
\hline
&84\\
&116\\  
&130\\  
&140\\  
$RL$&166\\ 
&171\\  
&233\\  
&248\\  
&265\\  
\end{tabular}
\end{ruledtabular}
\end{table}  

\section{Fitting parameters for the $A_{g}$ CDW modes}\label{Fitting_parameters_for_CDW_modes}  
\begin{figure}[!t] 
\begin{center}
\includegraphics[width=\columnwidth]{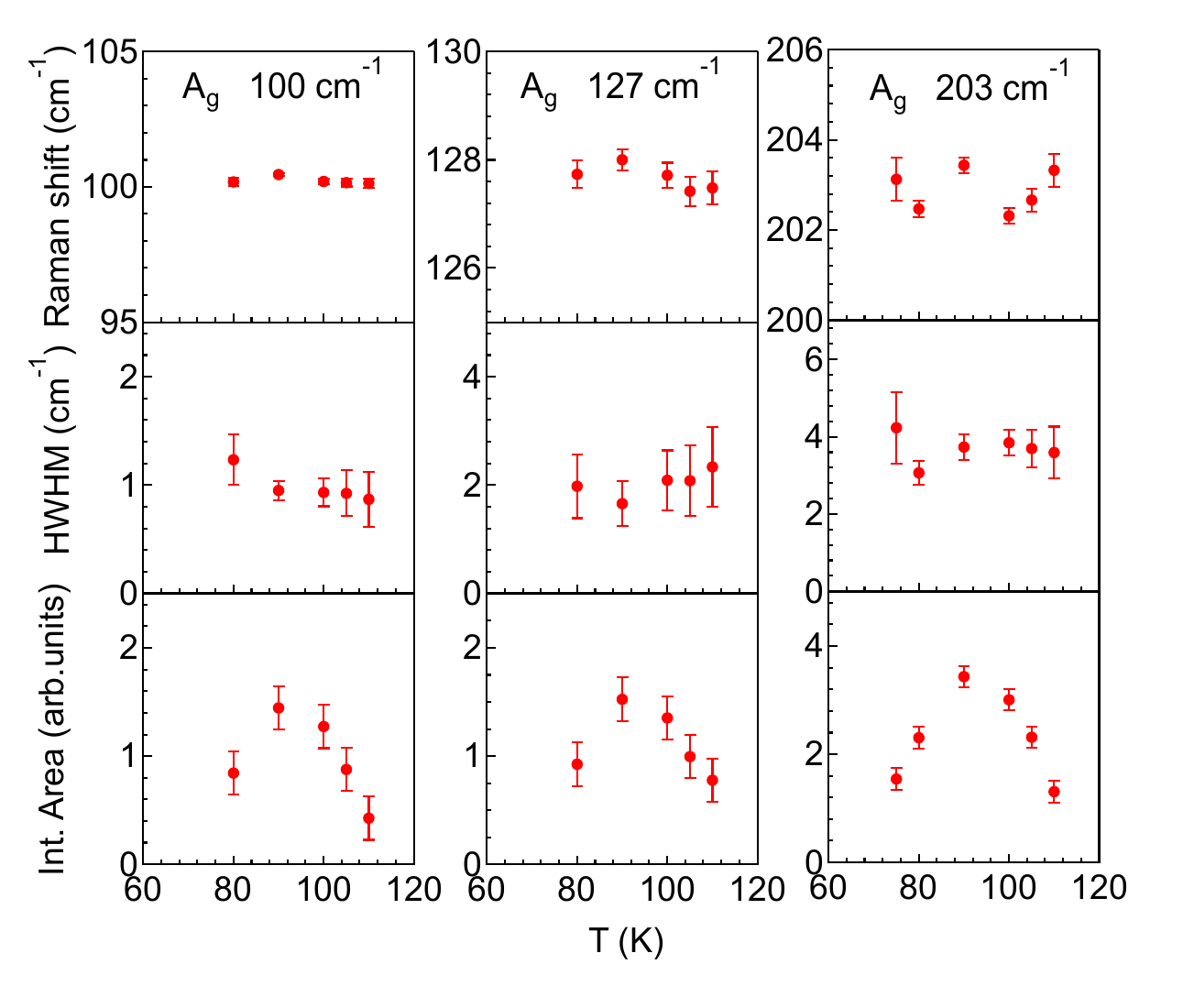}
\end{center}
\caption{\label{A1g_T_dependence} 
The $T$ dependence of the peak position, HWHM, and integrated intensity for the $A_{g}$ phonon mode at 100~\cm-1, 127~\cm-1, and 203~\cm-1 in the $RR$ scattering geometry.
The error bars represent one standard deviation.
}
\end{figure}

In this appendix, we discuss the $T$ dependence of the peak position, HWHM, and integrated intensity for the three $A_{g}$ phonon modes at 100~\cm-1, 127~\cm-1, and 203~\cm-1 in the $RR$ scattering geometries below $T_\text{CDW}$. As shown in Fig.~\ref{A1g_T_dependence},      
the $T$ dependence of the peak frequencies for the three modes does not change much between 60\,K and 110\,K. The HWHM for the three modes becomes a bit larger when approaching  $T_\text{canting}$, similar to the $A_{g}$ mode at 160~\cm-1 shown in Fig.~\ref{Fig4_T_dependence}(c) of the main text. The integrated intensities for the three modes first increase below $T_\text{CDW}$, reach a maximum at 90\,K, then decrease below 90\,K, and finally become zero at $T_\text{canting}$. They are consistent with the $T$ dependence for the $A_{g}$ mode at 160~\cm-1 shown in Fig.~\ref{Fig4_T_dependence}(c) of the main text.

\section{Coupled two-Lorentzian-phonon model}\label{Fitting_model_for_B1g_modes}  
In this appendix, we present the fitting model for the peak at 162\,\cm-1 and the shoulder peak between $T_\text{CDW}$ and $T_\text{canting}$ in the $RL$ scattering geometry. 

Both the peak at 162\,\cm-1 and the shoulder peak on the lower-energy side have $B_g$ symmetry; thus, they are allowed to be coupled.
 An interaction between the two modes can induce spectra weight transfer and also repelling of energy levels.
We use Green's function formalism to construct a model describing the main mode ($p_1$) and its shoulder peak ($p_2$)~\cite{Ye_2021_PhysRevB}. 

The Raman response of the coupled modes can be calculated from an interacting Green's function:
\begin{equation}\label{B1g_model_1}
\chi^{\prime \prime} \sim \operatorname{Im} T^{\text{T}} G T,
 \end{equation} 
where $T=[t_{p1}, t_{p2}]$, and $t_{p1}$ and $t_{p2}$ represent the light coupling amplitudes to the main peak $p_1$ and the shoulder peak $p_2$ in the $RL$ scattering geometries, respectively. The superscript $\text{T}$ denotes the transpose operation, and $G$ is the Green's function for the two interacting phononic systems. 
The Green's function $G$ can be obtained via the Dyson equation
\begin{equation}\label{B1g_model_2}
G=\left(G_0^{-1}-V\right)^{-1},
 \end{equation}                                                                                                                                                                                    
where $G_0$ is the bare Green's function and $V$ represents the interaction. 
Here, we consider two Lorentzian peaks that are coupled to each other. 
                                                        
The bare Green's function $G_0$ is
\begin{equation}\label{B1g_model_3}
G_{0}=\left(\begin{array}{cc}
G_{p1} & 0  \\
0 & G_{p2} \\
\end{array}\right),
 \end{equation}     
where $G_{p1}$ and $G_{p2}$ represent the bare main mode and the bare shoulder mode, respectively. 
They have the Lorentzian forms $G_{p1}=-1 /\left(\omega-\omega_{p1}+i \gamma_{p1}\right)+1 /\left(\omega+\omega_{p1}+i \gamma_{p1}\right)$ and $G_{p2}=-1 /\left(\omega-\omega_{p2}+i \gamma_{p2}\right)+1 /\left(\omega+\omega_{p2}+i \gamma_{p2}\right)$, where $\omega_{p1}$ and $\omega_{p2}$ are bare frequencies, and $\gamma_{p_1}$ and $\gamma_{p_2}$ are bare HWHM. 
                                      
Note that $V$ is an off-diagonal matrix that describes the coupling strength $v$ between the two modes:
\begin{equation}\label{B1g_model_4}
V=\left(\begin{array}{cc}
0 & v  \\
v & 0 \\
\end{array}\right).
 \end{equation} 
Inserting Eq.~(\ref{B1g_model_2})-(\ref{B1g_model_4}) to Eq.~(\ref{B1g_model_1}), we obtain the coupled two-Lorentzian phonon model. By fitting the Raman data shown in Fig.~\ref{Fig4_T_dependence}(e), we obtain the the bare mode energy, HWHM, and the light coupling amplitudes, as well as the integrated areas for the two bare modes between $T_\text{CDW}$ and $T_\text{canting}$ in the $RL$ scattering geometry. The temperature dependence of these parameters is shown in Fig.~\ref{Fig4_T_dependence}(f) of the main text. \\

\section{Anharmonic phonon decay model}\label{Anharmonic_decay_model}    
In this appendix, we discuss the anharmonic phonon decay model.
We fit the temperature dependence of the phonon frequency and HWHM by using the anharmonic phonon decay model~\cite{Klemens_PhysRev148,Cardona_PRB1984}: 
\begin{equation}
\label{eq_omega1}
\omega(T)=\omega_{0}-C_1\left[1+ 2n(\Omega(T)/2) \right],\\
\end{equation}
\begin{equation}
\label{eq_gamma1}
\Gamma(T)=\gamma_{0}+\gamma_1\left[1+ 2n(\Omega(T)/2)\right],
\end{equation}
where $\Omega(T)= \hbar \omega / k_BT$, $n(x)=1/(e^x-1)$ is the Bose-Einstein distribution function. Note that $\omega(T)$ and $\Gamma(T)$ involve mainly three-phonon decay processes where an optical phonon decays into two acoustic modes with equal energy and opposite momentum. 


%


\end{document}